\documentclass[pra,aps,showpacs,footinbib,twocolumn]{revtex4-1}
\usepackage{amssymb}
\usepackage{graphicx}
\usepackage{amsmath}
\usepackage{times}
\usepackage{color}
\usepackage{setspace}
\usepackage{bm}
\usepackage{float}
\usepackage{subfigure}
\usepackage[colorlinks, linkcolor=blue, citecolor=blue, urlcolor=blue, breaklinks=true]{hyperref}
\usepackage{braket}
\usepackage{mathtools}
\usepackage{placeins}
\usepackage{gensymb}
\usepackage{bm}
\usepackage{ctable}
\usepackage{braket}
\usepackage{epstopdf}
\usepackage{mathrsfs}

\DeclareFontFamily{OT1}{pzc}{}
\DeclareFontShape{OT1}{pzc}{m}{it}{<-> s * [1.10] pzcmi7t}{}
\DeclareMathAlphabet{\mathpzc}{OT1}{pzc}{m}{it}
\allowdisplaybreaks

\begin{document}

\title{Light scattering from dense cold atomic media}

\author{Bihui Zhu}
\affiliation{JILA, NIST and University of Colorado, 440 UCB, Boulder, CO 80309, USA}
\affiliation{Department of Physics, University of Colorado, 440 UCB, Boulder, CO 80309, USA}

\author{John Cooper}
\affiliation{JILA, NIST and University of Colorado, 440 UCB, Boulder, CO 80309, USA}
\affiliation{Department of Physics, University of Colorado, 440 UCB, Boulder, CO 80309, USA}
\author{Jun Ye}
\affiliation{JILA, NIST and University of Colorado, 440 UCB, Boulder, CO 80309, USA}
\affiliation{Department of Physics, University of Colorado, 440 UCB, Boulder, CO 80309, USA}
\author{Ana Maria Rey}
\affiliation{JILA, NIST and University of Colorado, 440 UCB, Boulder, CO 80309, USA}
\affiliation{Department of Physics, University of Colorado, 440 UCB, Boulder, CO 80309, USA}
\date{\today}

\date{\today}

\begin{abstract}
 We theoretically study the propagation of light through a cold atomic medium, where the effects of motion,  laser intensity, atomic density, and polarization can all modify the properties of the scattered light.  We present two different microscopic models: the ``coherent dipole model'' and the ``random walk model'',  both suitable for modeling recent  experimental work done in large atomic arrays in the low light intensity regime. We use them to compute relevant observables such as   the linewidth, peak intensity and line center of the emitted light. We  further develop generalized models that explicitly take into account  atomic motion. Those are relevant for hotter atoms and   beyond the low intensity regime. We show that atomic motion can lead to drastic dephasing and to a reduction of collective effects, together with a distortion of  the lineshape. Our results are applicable to model a full gamut of quantum systems that rely on  atom-light interactions  including atomic clocks, quantum simulators and  nanophotonic systems.
\end{abstract}

\maketitle
\section{Introduction}
Light-matter interactions are  fundamental for the control and manipulation of quantum systems. Thoroughly understanding them can lead to significant  advancements in  quantum technologies,  quantum simulations, quantum information processing and precision  measurements   \cite{kimblenature2008,ciracprl1997,solid2008,douglas2015,bloch2005,Haldane2014,bloch2012}. Over the past decades, cold atom experiments have provided a clean and tunable platform for studying light-matter interactions in microscopic systems where rich quantum effects emerge, such as superradiance and subradiance, electromagnetically induced transparency, and non-classical states of light \cite{bohnet2012,subradiance2015,EITnature2011,nonclassical2011,nonclassical2012}. However, in spite of intensive theoretical and experimental efforts over the years, long-standing open questions still remain regarding the propagation of light through a coherent medium, especially when it consists of  large  and dense ensembles of scatterers \cite{dfvjames,lehmberg,jannePRA1997,juhapra1999,scully2006,scully,subradiance,sokolov2013scaling,juhashift,sokolov2015,havey2015,antoine2015,juha2016,antoine2016,kaiser2016superrad}. In fact, by studying small systems where analytical solutions are obtainable, it has been realized that  atom-atom interactions can significantly modify the spectral characteristics of the emitted light. These effects yet need to be understood in large systems \cite{igor2013,ritsch2014,yu2015,zhu2015,bettles2015} where finite size effects and boundary conditions become irrelevant. The situation is even more  complicated when the coupling with atomic motion is non-negligible \cite{berman1997,li2013recoil}. It is timely to develop  theories capable of addressing these questions, given  the rapid developments on  cold atom experiments and nanophotonic systems. The experiments are entering strongly coupled regimes, where atom-atom and atom-photon interactions need to be treated simultaneously and sometimes fully microscopically \cite{igor2013,antoine2015b,goldschmidt2015}.

A widely adopted approach to describe light scattering  consists of integrating out the atomic degrees of freedom and treating the atoms just as random scatterers with prescribed polarizability \cite{scatreview1998,scatreview1999,mcrobin1,scatterbook}. While this approach can successfully capture some classical properties of the scattered light, it does not fully treat the roles of atom-atom interactions and atomic motion \cite{kaiserprl1999,labeyrie2004multiple,kaiserprl2004,thermal}. An alternative route consists of tracing over  the photonic degrees of freedom. In this case the virtual exchange of photons induces dispersive and dissipative dipole-dipole interactions  between atoms,  which can be accounted for by  a master equation formulation \cite{dfvjames,lehmberg}. This approach has been used to study  systems of tightly localized atoms where the dynamics only takes place in the atomic  internal degrees of freedom. It has been shown to successfully capture quantum effects in light scattering  \cite{scully,dalibard,yu2014,subradiance,opticaltheo,purdue}. However, due to the computational complexity, it has been often restricted to weak excitation and  small samples \cite{klaus2013,meir2014,yu2015,zhu2015}, and a direct comparison with   experiments containing a large number of atoms has been accomplished only recently \cite{bromley2016,havey2016exp}. In general, most theories have  not properly accounted for atomic motional effects  and atomic  interactions on the same footing  and many open questions  in light scattering processes remain.

\begin{figure}
\centering
\includegraphics[width=0.4\textwidth]{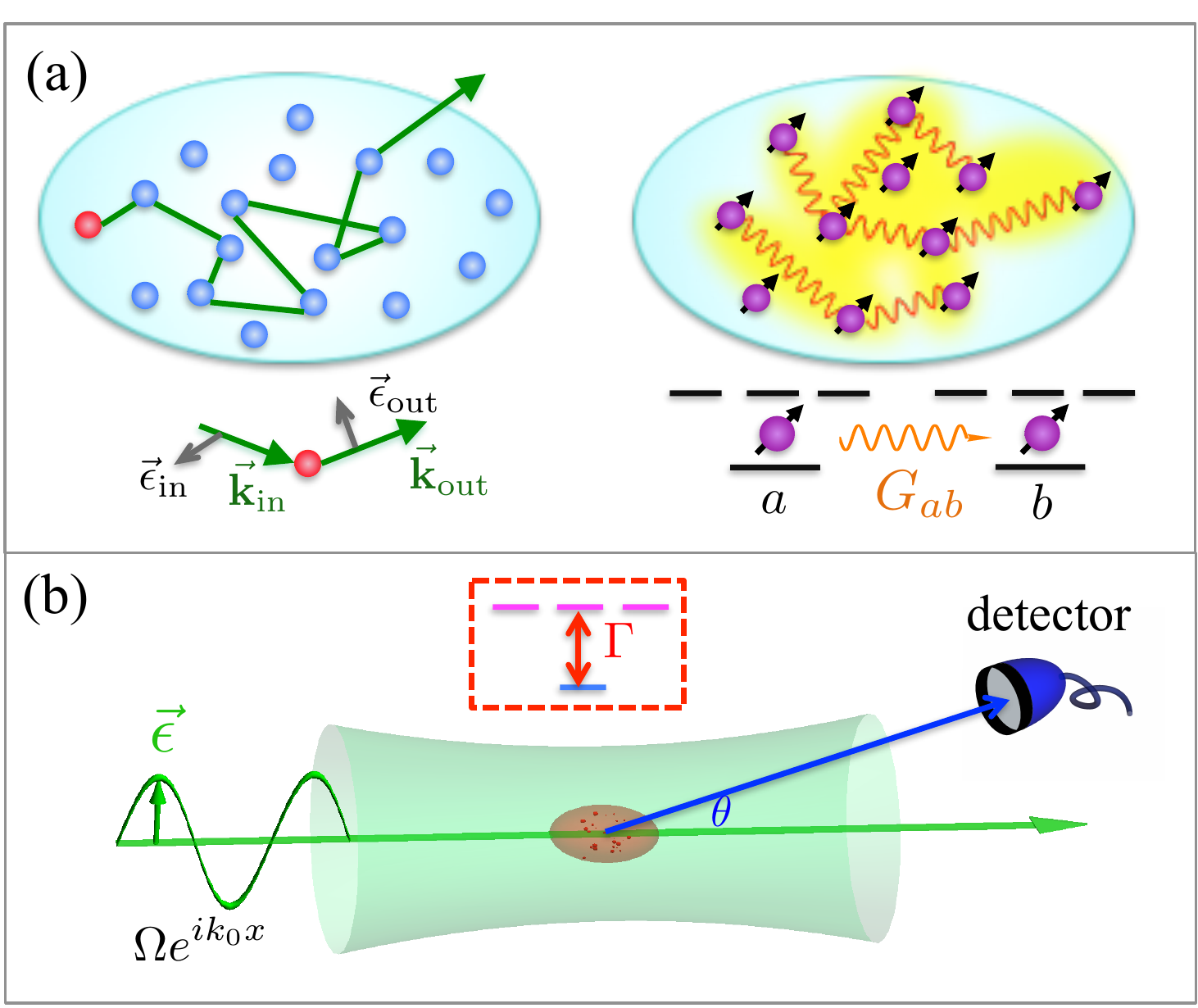}
  \caption{(Color online) {\bf Scheme}. (a) Microscopic models:   random walk model (left) and coherent dipole (right). In the random walk model, a photon is  randomly scattered by the atoms. Scattering events are characterized by the incident and outgoing wavevector of the photons and their corresponding polarization. In the coherent dipole model, atoms are coupled by dipole-dipole interactions ($G_{ab}$) and  all  atoms contribute to the fluorescence.  (b)  Experimental setup for measuring the fluorescence from a cloud of atoms. An incident laser drives an atomic transition with a spontaneous emission  rate $\Gamma$. Atoms absorb and emit light. The detectors collect   scattered photons at an angle $\theta$ measured from  the incident beam direction.  }\label{fig:scheme}
\end{figure}

Here, we present a unifying theoretical framework based on a coherent dipole (CD) model   (see Fig.~\ref{fig:scheme}(a)) to study the light scattering from cold atoms with possible residual motion. In the low intensity and slow motion regime, we use the CD to investigate the collective effects in the light scattered by a large cloud and show the interplay of optical depth and density. These results are compared with the random walk (RW) model   (see Fig.~\ref{fig:scheme}(a)) that only accounts for incoherent scattering and thus ignores coherent dipolar interaction effects. To address the role of atomic motion, we perform different levels of generalization of the CD model. With these modified models, we show that  atomic motion not only reduces phase coherence and collective effects, but also  impacts the lineshape and linecenter of the spectral emission lines via photon recoils. Motivated by a recent experiment at JILA (see Fig.~\ref{fig:scheme}(b)) \cite{bromley2016}, we focus our discussions on a $J=0\rightarrow J=1$  transition, but the methods presented here can be extended to more complicated level structures without much difficulty.

This paper is organized as follows. In Sec. II, we provide the mathematical description of the CD model, which treats atoms as coupled, spatially fixed dipoles sharing a single excitation. Its predictions on the collective properties of the emitted light
such as the light polarization  and density dependence  of the  lineshape and peak intensity  are discussed. In Sec. III,  we introduce the RW model and  compare its predictions on the linewidth and peak intensity of the scattered light to the ones obtained from the CD model. Those comparisons allow us to explore the role of   phase coherence in the atom-light interaction. In Sec. IV, we present extended models to study  motional effects on atomic emission. We first include motion in the CD model by assuming that its leading contributions comes from  Doppler shifts. Those are accounted for in the frozen model approximation when we introduce local random detunings for atoms sampled from  a Maxwell-Boltzmann distribution. Then we go beyond the frozen model approximation and explicitly include atomic motion  by means of  a semi-classical approach. This treatment also allow us to go  beyond the low excitation regime.  We finish in  Sec. V with conclusions and and outlook.   

\section{Coherent dipole model}\label{sec:CD}
\subsection{Equations of motion for coherent dipoles}
For an ensemble of $N$ atoms with internal dipole transition $J=0\rightarrow J'=1$, the Hamiltonian of the system that includes  the interaction between atoms and the radiation field is \cite{haroch1982}
\begin{eqnarray}
H&=&\hbar\sum_{{\bf k},\boldsymbol{\epsilon}}\omega_k\hat{a}^\dagger_{{\bf k}\boldsymbol{\epsilon}}\hat{a}_{{\bf k}\boldsymbol{\epsilon}}+\hbar\sum_{i,\alpha}\omega_\alpha\hat{b}_{i}^{\alpha\dagger}\hat{b}_i^{\alpha}+\sum_i\hat{\bf D}_i\cdot\hat{\bf E }({\bf r}),\nonumber\\
\\
\hat{D}_i^\alpha&=&d(\hat{b}_i^{\alpha\dagger}+\hat{b}_i^{\alpha}),\\
\hat{\bf E}({\bf r})&=&\sum_{{\bf k}{\boldsymbol{\epsilon}}}g_k\boldsymbol{\epsilon}_{\bf k}(e^{i{\bf k}\cdot{\bf r}}\hat{a}_{{\bf k}\boldsymbol{\epsilon}}+\mathrm{H.c.}),
\end{eqnarray}
where  we have used the notation $\ket{\alpha}$ to denote the excited levels, and $\ket{0}$ for the lower state. For convenience, we  choose the Cartesian basis, $\ket{\alpha}=\ket{x}, \ket{y}$ or $\ket{z}$. $\hat{b}_i^{\alpha\dagger}=\ket{\alpha_i}\bra{0_i}$ is the raising operator for transition to state $\ket{\alpha}$ of the $i^{\mathrm{th}}$ atom, and $d$ is the atomic dipole moment. The field coupling strength is denoted by $g_k=\sqrt{\frac{\hbar\omega_k}{2\varepsilon_0V}}$,   ${\bf k}~(\boldsymbol{\epsilon})$ is the wavevector (polarization) of the photons, $\omega_k$ is the frequency of the photons, $\varepsilon_0$ is the vacuum permitivity, and $V$ is the photon quantization volume. Under the Born-Markov approximation, the photon degrees of freedom,  $\hat{a}_{\bf k\epsilon}$, can be adiabatically eliminated, leading to a master equation for the reduced density matrix ($\hat{\rho}$) of the  atoms \cite{haroch1982,dfvjames,igor2013}  where the effective role of the  scattered photons is to mediate  dipole-dipole interactions between atoms. The master equation for $\hat{\rho}$ is
\begin{eqnarray}
i\frac{d\hat\rho}{dt}&=&-\sum
_{i,\alpha}\Delta^\alpha\ket{\alpha_i}\bra{\alpha_i}+\sum_{i,\alpha}\Omega^\alpha(e^{i{\bf k}_0\cdot{\bf r}_i}\hat{b}_i^{\alpha\dagger}+\mathrm{H.c.})\nonumber\\&& +\sum_{i\neq j,\alpha,\alpha'}g_{i,j}^{\alpha\alpha'}[\hat{b}_i^{\alpha\dagger}\hat{b}_j^{\alpha'},\hat{\rho}]
+\sum_{i,j,\alpha,\alpha'}f_{i,j}^{\alpha\alpha'}[[\hat{b}_i^{\alpha\dagger},\hat\rho],\hat{b}_j^{\alpha'}],\nonumber\\\label{eq:master}
\end{eqnarray} 
where we have added the term describing the effect of an external driving laser with polarization $\alpha$, wavevector ${\bf k}_0$, and Rabi frequency $\Omega^\alpha$. The Hamiltonian is written in   the rotating frame of the laser, with $\Delta^\alpha$ denoting the detuning between the laser and the transition $\ket{0}\rightarrow\ket{\alpha}$.  The dipole-dipole interactions are given by \cite{dfvjames,igor2013}
\begin{eqnarray}
G_{ij}^{\alpha\alpha'}&=&\frac{3\Gamma}{4}[\delta_{\alpha,\alpha'}A(r_{ij})+\hat{\bf r}_{ij}^\alpha\hat{\bf r}_{ij}^{\alpha'}B(r_{ij})],\\
A(r)&=&-\frac{e^{ik_0r}}{k_0r}-i\frac{e^{ik_0r}}{k_0^2r^2}+\frac{e^{ik_0r}}{k_0^3r^3},\\
B(r)&=&\frac{e^{ik_0r}}{k_0r}+3i\frac{e^{ik_0r}}{k_0^2r^2}-3\frac{e^{ik_0r}}{k_0^3r^3},\\
g_{ij}^{\alpha\alpha'}&=&\Re[G_{ij}^{\alpha\alpha'}],\\
f_{ij}^{\alpha\alpha'}&=&\Im[G_{ij}^{\alpha\alpha'}],\label{eq:dipole}
\end{eqnarray} 
where $\delta_{\alpha,\alpha'}$ is the Kronecker delta symbol, ${\bf r}_{ij}$ is  the relative seperation  between the  atoms $i$ and $j$,   and $\hat{\bf r}^\alpha={\bf r}^\alpha/r$ denotes the component of the unit vector ${\bf r}/r$ along the direction $\alpha=x,y$ or $z$. The real  and imaginary parts describe the dispersive and dissipative interactions, respectively. The spontaneous emission rate is $\Gamma=\frac{k_0^3d^2}{3\pi\hbar\varepsilon_0}$, and $k_0=2\pi/\lambda$ is the wavevector of the dipole transition. The dipole-dipole interactions include both the far-field ($1/r$) and near-field ($1/r^2,1/r^3$) contributions. The imaginary part encapsulates collective dissipative process responsible for the superradiant emission in a dense sample. The real part accounts for elastic interactions between atoms which can give rise to coherent dynamical evolution. These elastic interactions compete  with and can even  destroy the superradiant emission \cite{friedberg1972,zhu2015}. 

When the atoms' thermal velocity $v$ satisfies $k_0v\ll\Gamma$, atoms can be assumed to be  frozen during the  radiation process. Moreover, in the weak driving regime, $\Omega\ll\Gamma$, to an excellent approximation, the master equation dynamics can be captured by  the  $3N$ linear equations describing the atomic coherences  $b_j^\alpha=\langle\hat{b}_j^{\alpha}\rangle$ of  an excitation propagating through the  ground state atomic medium.  The corresponding steady state solution  can be found from:
\begin{eqnarray}
b_j^\alpha&=&\frac{\Omega^\alpha\delta_{\alpha,\gamma} e^{i{\bf k}_0\cdot{\bf r}_j}/2}{\Delta^\alpha+i\Gamma/2}+\sum_{n\neq j,\alpha'}\frac{G_{jn}^{\alpha\alpha'}}{\Delta^\alpha+i\Gamma/2}b_n^{\alpha'},\label{eq:steady}
\end{eqnarray} where we have specified the polarization of the driving laser to be along $\gamma$.
The fluorescence intensity measured  at the position  ${\bf r}_s$ in the far-field can be obtained by the summation \cite{dfvjames}
\begin{eqnarray}
I({\bf r}_s)\propto\sum_{jn}e^{-i{\bf k}_s\cdot{\bf r}_{jn}}\sum_{\alpha,\alpha'}(\delta_{\alpha,\alpha'}-\hat{\bf r}_s^\alpha\hat{\bf r}_s^{\alpha'})b_j^{\alpha'}b_n^{\alpha*},\label{eq:int}
\end{eqnarray}
where $b_n^{\alpha*}$ is the complex conjugate of $b_n^{\alpha}$, and ${\bf r}_{jn}={\bf r}_j-{\bf r}_n$.
\subsection{Collective effects in fluorescence}\label{subsec:colint}
For dilute samples  the dipolar interactions are weak, $\mathcal{G}\equiv \sum_{i\neq j,\alpha,\alpha'}|G^{\alpha\alpha'}_{ij}|/(N\Gamma)\ll 1$,  and  Eq. (\ref{eq:steady}) can be solved perturbatively using   $\mathcal{G}$ as an  the expansion parameter, $b_j^\alpha=b_j^{\alpha,0}+b_j^{\alpha,1}+b_j^{\alpha,2}+...$, ($b_j^{\alpha,n}\propto \mathcal{G}^n$) which results in
\begin{eqnarray}
b_j^{\alpha,n}&=&\sum_{\mathclap{\substack{l_1\neq j\\l_2\neq l_1\\...l_n\neq l_{n-1}\\\alpha_1,\alpha_2...\alpha_n}}}\frac{G_{jl_{1}}^{\alpha\alpha_1}G_{l_1l_{2}}^{\alpha_1\alpha_2}...G_{l_{n-1}l_{n}}^{\alpha_{n-1}\alpha_n}\Omega^\gamma\delta_{\alpha_n,\gamma}e^{i{\bf k}_0\cdot{\bf r}_{l_n}}}{i^n(\Delta^\alpha+i\frac{\Gamma}{2})(\Delta^{\alpha_1}+i\frac{\Gamma}{2})...(\Delta^{\alpha_n}+i\frac{\Gamma}{2})}.\label{eq:expand}
\end{eqnarray} 
In the expansion, terms of order $n$ account for $n ^{\rm th}$ order scattering events. For simplicity, in the following we assume the atomic sample has a spherical shape, with density distribution $n(r)=n_0 e^{-r^2/2R^2}$, unless otherwise specified. However, the conclusions  can be generalized to other geometries. Here, $n_0=\frac{N}{2\pi\sqrt{2\pi}R^3}$ is the peak density.

To the zeroth order, the atomic response is driven by the external field only and is not modified by the scattered light:
\begin{eqnarray}
b_j^{\alpha,0}=\frac{\Omega^\alpha\delta_{\alpha,\gamma}e^{i{\bf k}_0\cdot{\bf r}_j}}{\Delta^\alpha+i\Gamma/2}.\label{eq:0th}
\end{eqnarray}
Substituting it into Eq. (\ref{eq:int}), the intensity of scattered light is
\begin{eqnarray}
I\propto N +N^2e^{-|{\bf k}_s-{\bf k}_0|^2R^2}.
\end{eqnarray}
There are two contributions to the intensity, the first term $\propto N$ represents the incoherent contribution, and the second term $\propto N^2$ is the collective emission resulting from  coherent scattering processes\cite{scully2006,bromley2016}. The phase coherence is restricted to a narrow angular region around  the incident laser direction, with $\delta\theta\sim1/k_0R$. The enhanced emission arises  from the constructive interference of the radiation from $N$ dipoles \cite{opticaltheo}. Along other directions, the  random distribution of atom positions randomizes the phases of the emitted light, smearing out the phase coherence after averaging over the whole sample \cite{bromley2016}.

Including first order corrections,  the intensity of the scattered light is given by 
\begin{eqnarray}
I(r_s)&\propto&\frac{N\Omega^2}{(\Delta-\Gamma\Re[\overline{G}])^2+(\Gamma+2\Gamma\Im[\overline{G}])^2/4},\label{eq:1st}
\end{eqnarray}
 for transverse directions, where we have denoted $\Omega=\Omega^\gamma$ and  $\Delta=\Delta^\gamma$, and $\overline{G}=\sum_{i\neq j}G_{ij}^{\gamma\gamma}e^{-i{\bf k}_0\cdot{{\bf r}_{ij}}}/(N\Gamma)$. For the forward direction, the intensity has the same form, except that the factor $N$ is replaced by $N^2$, due to the  phase coherence. Therefore, the lineshape of the scattered light is Lorentzian, with its line center
 frequency shifted by the elastic interactions, and  the linewidth broadened by the radiative interactions.  If we temporarily neglect the effect of polarization, when the atom-atom separation is large, the dipole-dipole interaction is dominated by far-field terms, {\em i.~e.}, $G_{ij}^{\gamma\gamma}\sim-\frac{3\Gamma}{4}(1-\hat{\bf r}_{ij}^\gamma \hat{\bf r}_{ij}^\gamma)\frac{e^{ik_0r_{ij}}}{k_0r_{ij}}$. In this limit analytical expressions for the linewidth broadening $\overline\Gamma$, and density shift, $\overline\Delta$, can be obtained: $\overline\Gamma=2\Gamma\Im[\overline{G}]=3N\Gamma/(8k_0^2R^2)= \frac{\rm OD}{4}\Gamma$, with ${\rm OD}=\frac{3N}{2k_0^2R^2}$ the optical depth of the sample (see Appendix \ref{app:od}), and  $\overline\Delta=\Gamma\Re[\overline{G}]=-\Gamma n_0k_0^{-3}/4\sqrt{2\pi}$. Therefore, while the collectively broadened linewidth depends on the OD of the atomic cloud, the frequency shift depends on the density. 

\begin{figure*}
\centering
\includegraphics[width=0.7\textwidth]{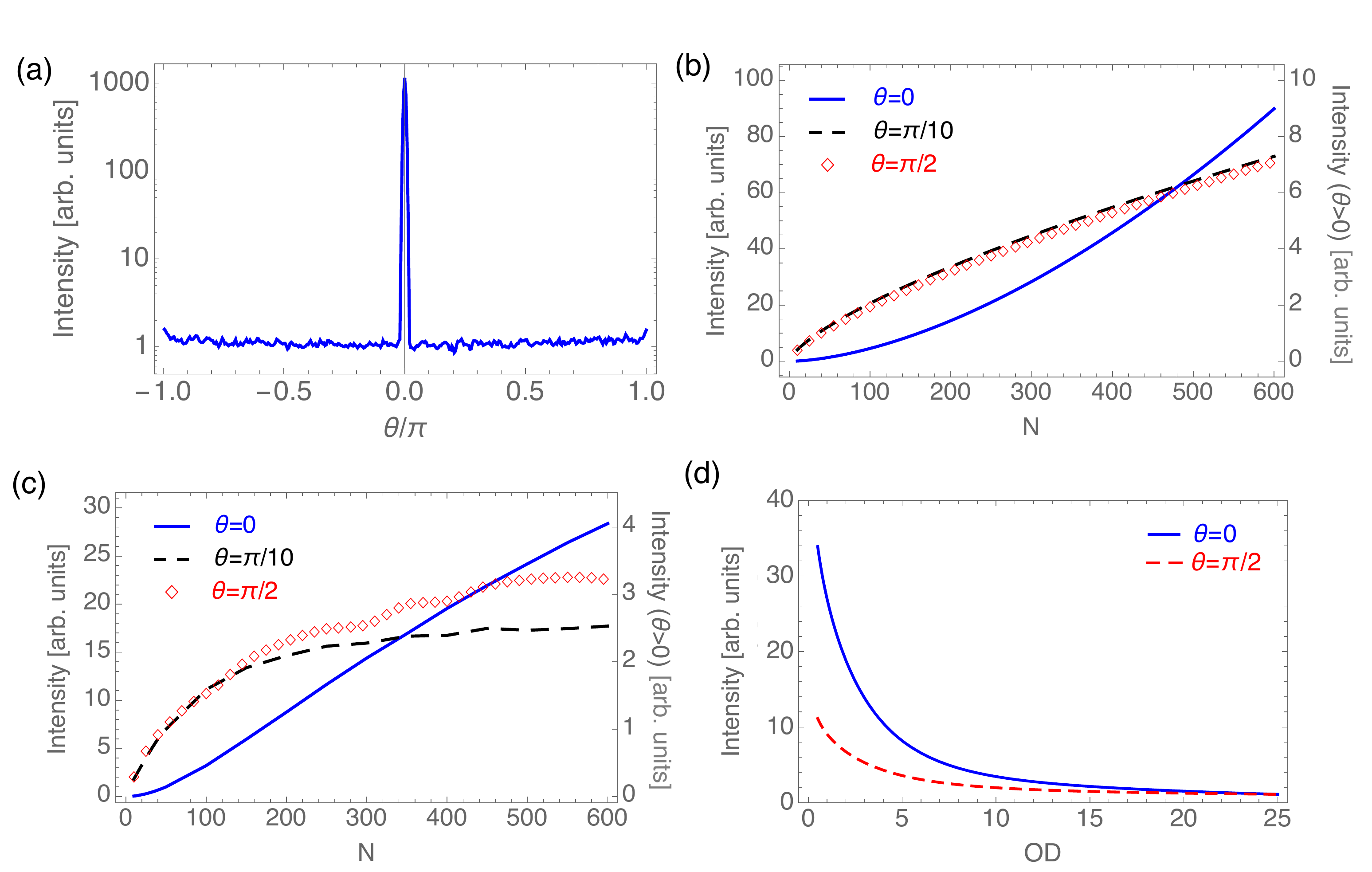}
  \caption{(Color online) {\bf CD Model: on-resonance  intensity}.  (a) Angular distribution: due to the constructive interference along the forward direction ($\theta=0$), the intensity is drastically enhanced within a small angular region. (b)-(c) The  intensities for three different directions ($\theta=0,\pi/10,\pi/2$) are shown as a function of atom number, each normalized to the value at $N=50$. All are calculated  for a spherical cloud of fixed size. The right vertical axes label the intensity for $\theta=\pi/2,\pi/10$. In (b) the  OD and the density are relatively low (when $N=500$, OD=2,$n_0k_0^{-3}$=0.0015). The transverse intensity increases $\sim N$, while the forward intensity increases $\sim N^2$, showing  a collective enhancement for small window of $\theta$ around zero. Outside this narrow angular window the enhancement disappears and the intensity becomes almost $\theta$ independent as   indicated by the identical behavior  observed for two different angles, $\theta=\pi/10$ and $\pi/2$.  In (c) the OD and density are relatively large (when $N=500$, OD=10, $n_0k_0^{-3}$=0.017). The rate at which the intensity increases with $N$ slows down in both the forward and transverse direction: with $I\sim N^1$ for $\theta=0$ and $I\sim N^{0.5}$ for $\theta=\pi/2$, respectively. (d) On-resonance intensity as a function of OD: it is highly suppresed at large OD. Here, the intensity is normalized to the corresponding value at ${\rm OD}=25$ for each direction. }\label{fig:multipscat}
\end{figure*} 

\begin{figure}
\centering
\includegraphics[width=0.35\textwidth]{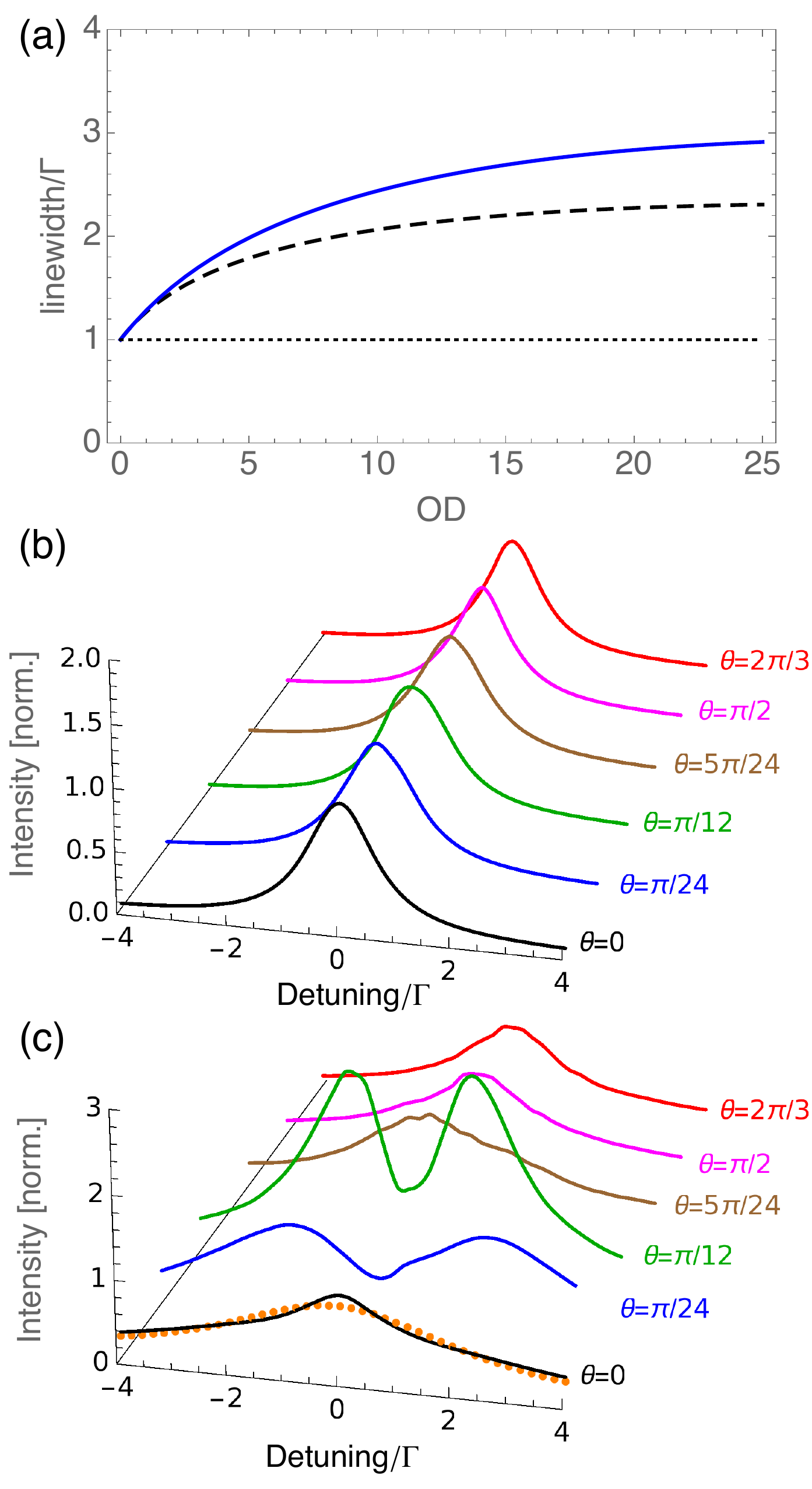}
  \caption{(Color online) {\bf CD Model: collective broadening}.   (a) FWHM linewidth as a function of OD. At small OD, the FWHM increases linearly with OD, but as OD increases density effects set in, multiple scattering events become relevant and the linewidth dependence on OD is no longer linear. Two different atom numbers are used for the blue (N=1000) and black lines (N=200), and the OD is varied by changing the cloud size. For the blue line,  $n_0k_0^{-3}=0.14$ at OD=25, $0.003$ at OD=2.  For the black line, the density at the same OD is doubled. With smaller density, the linewidth increases to a larger value in the large OD regime. (b) The lineshape at small OD values (OD=2, $n_0k_0^{-3}=0.002$) for different angles $\theta$ ($\theta$ is defined in Fig.~\ref{fig:scheme}(a)) is mainly  Lorentzian. Here the intensity is normalized to the on-resonance intensity for each $\theta$.  (c) At large OD (OD=20, $n_0k_0^{-3}=0.06$), the fluorescence lineshape  significantly broadens and stops being  Lorentzian. The brown dots for $\theta=0$ show the Lorentzian fit, which fails to describe the lineshape. At intermediate $\theta$, a double-peak structure shows up. For all panels, the cloud shape is spherical.  }\label{fig:lwcol}
\end{figure}

 In a dense medium dipolar interactions are strong, $\mathcal{G}\gtrsim 1$, higher order scattering events become important, and  the interplay between the radiative interactions and elastic interactions becomes  non-negligible. As a consequence  the above perturbative analysis is no longer applicable.   In Fig.~\ref{fig:multipscat} and Fig.~\ref{fig:lwcol}, we show the numerical solution of Eq. (\ref{eq:steady}), which takes into account all the scattering orders. As shown in Fig.~\ref{fig:multipscat}(a), most of the scattered light is distributed within a narrow peak around the forward direction (laser direction). Outside this narrow region fluorescence is almost uniformly distributed among other  directions \cite{fnote1}. The forward emission is collectively enhanced. For low OD it increases  as $\sim N^2$ (Fig.~\ref{fig:multipscat} (b)) while the transverse intensity  increases as $\sim N$.

Dipolar interactions tend to suppress the rate at which the  intensity grows with $N$ (Fig.~\ref{fig:multipscat}(c)). This  can be qualitatively  understood from Eq. (\ref{eq:1st}), which predicts that the intensity is reduced as OD increases. Despite the fact that the perturbative analysis is only valid in the weak interaction limit, this tendency  remains and becomes more pronounced in  the large OD regime as shown by the numerical solution presented  in Fig.~\ref{fig:multipscat}(d).   Broadly speaking, multiple scattering events tend to suppress collective behavior \cite{juhashift,resonantmultiple}. Similar physics  is also  observed in the behavior of the linewidth.  At small OD,  the FWHM linewidth increases  linearly with OD (Fig.~\ref{fig:lwcol}(a)) and the lineshape is well described by a Lorentzian (Fig.~\ref{fig:lwcol}(b)), as expected from Eq. (\ref{eq:1st}). However, when OD is large and the density is high, in addition to a significant broadening, the lineshape becomes non-Lorentzian  (Fig.~\ref{fig:lwcol}(c)) and the FWHM  increases slowly with OD (Fig.~\ref{fig:lwcol}(a)). To further illustrate this, in Fig.~\ref{fig:lwcol}(a) we plot the FWHM for the same OD but with smaller density (by using a larger atom number). The figure shows that the linewidth indeed keeps increasing  until saturation at a larger value of OD. Another interesting feature is the double-peak structure for intermediate angles $\theta$, arising due to the competition between interference and multiple scattering events (Fig.~\ref{fig:lwcol}(c)).

\begin{figure}[tb]
\centering
\includegraphics[width=0.35\textwidth]{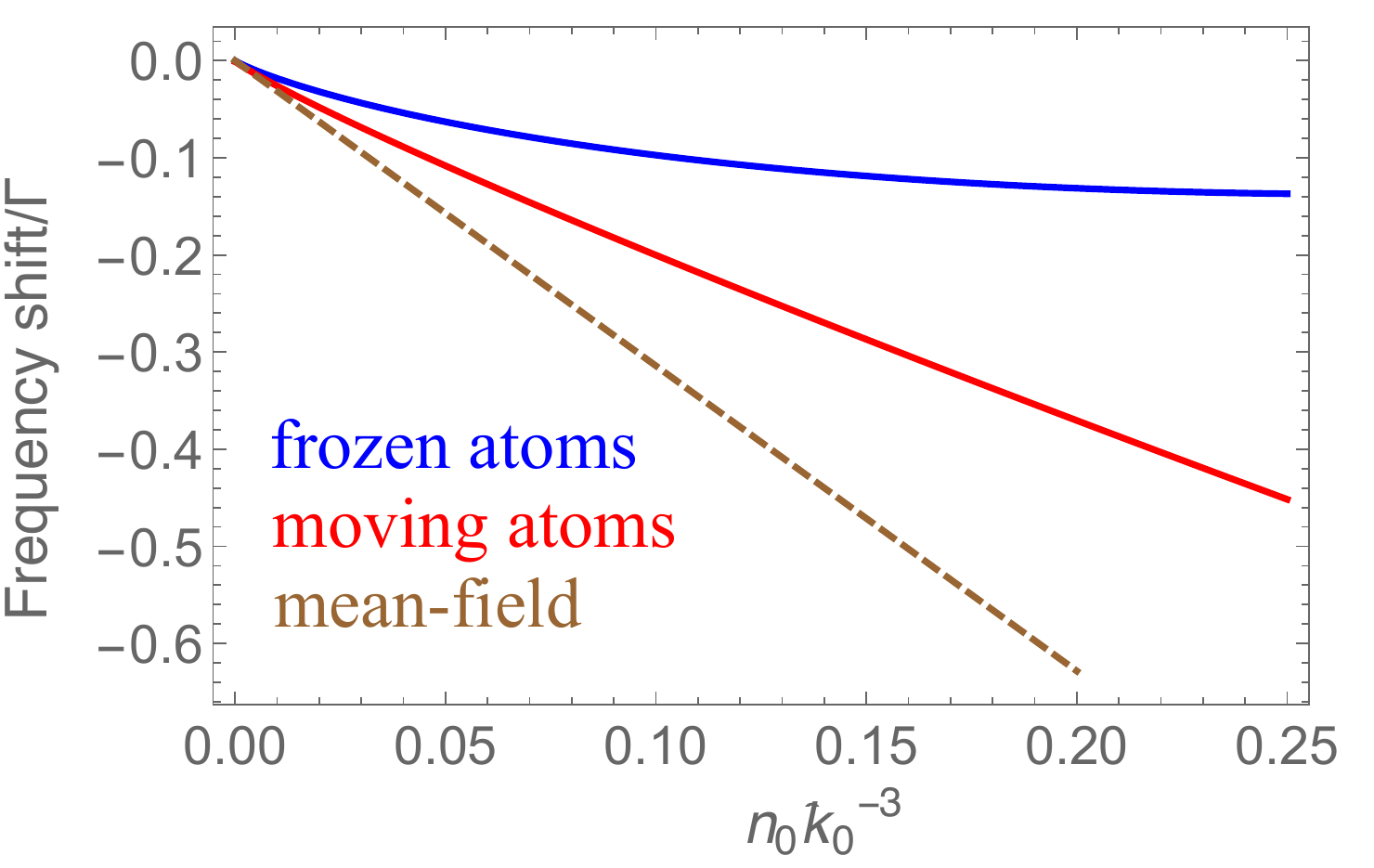}
  \caption{Color online) {\bf CD Model: frequency shift}. For small density, the shift calculated from the coherent dipole model (blue line) increases linearly with density as predicted by the mean-field theory. However, when density is large, there is a significant deviation from the mean-field result. When motional effects are taken into account (red line, Doppler width $=5\Gamma$, see Sec. \ref{subsec:modified}), the non linear suppression of  the frequency shift with density is less severe. }\label{fig:shifdens}
\end{figure}
The drastic modifications of the perturbative expectations from multiple scattering are also present in the  frequency shift of the scattered light.  From a mean-field point of view, the linecenter of scattered light is shifted according to the Lorentz-Lorenz shift $\pi n_0k_0^{-3}\Gamma$ \cite{freqshift}. As shown by the numerical calculation in Fig.~\ref{fig:shifdens} (blue line), at small density, the frequency shift is indeed linear with $n_0k_0^{-3}$, but as density increases, the shift is quickly suppressed \cite{juhashift}.  For  atom density $\sim 5\times 10^{13}$cm$^{-3}$ (reached for example  in cold $^{87}$Rb atom experiments \cite{haveypra2013}), the calculated density shift  is  approximately a factor of two lower  than the mean-field prediction.
\begin{figure}
\centering
\includegraphics[width=0.35\textwidth]{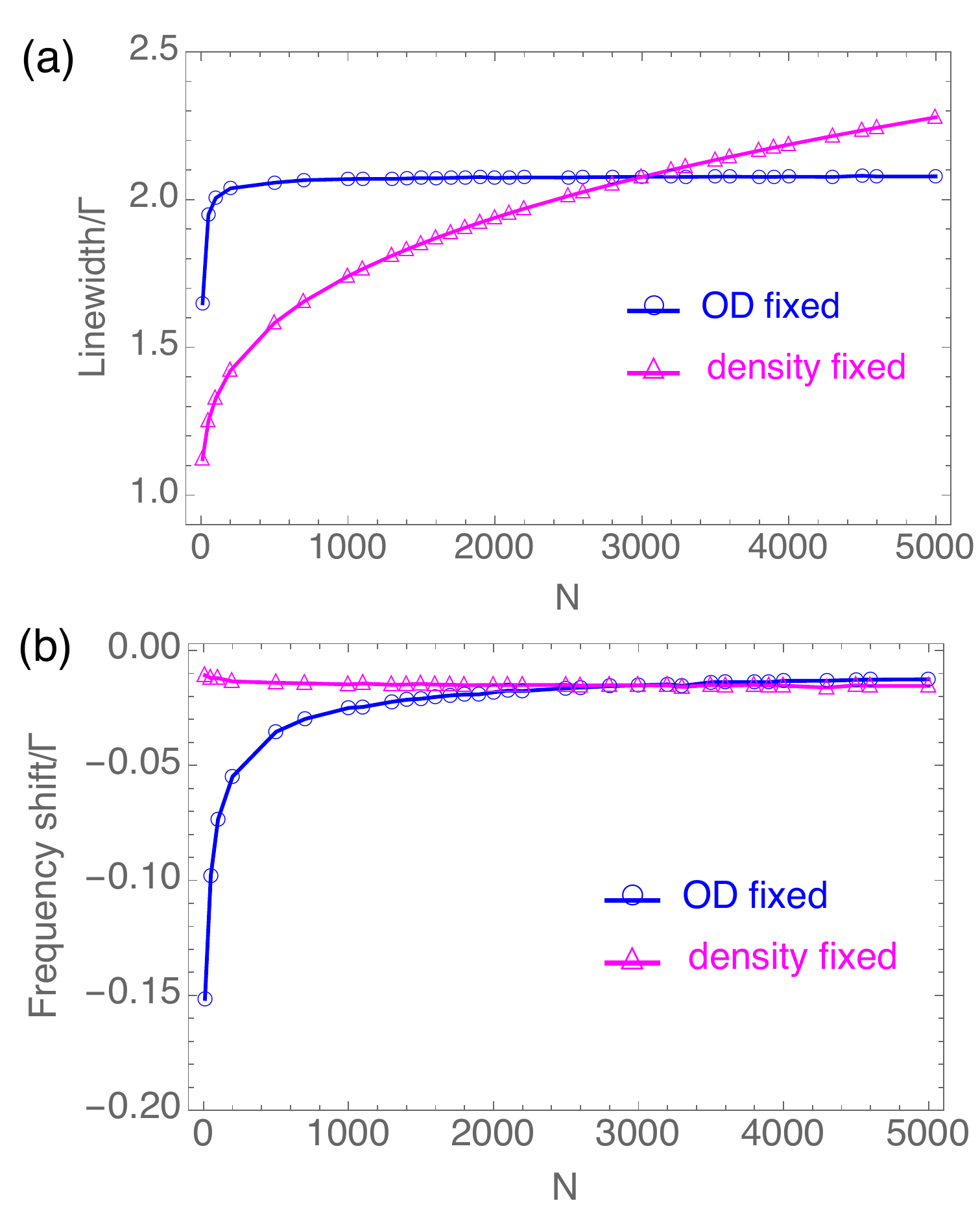}
  \caption{(Color online) {\bf CD Model: finite size scaling}. (a) The linewidth is calculated for different number of atoms at the same OD (OD=4 for $N$=3000) by varying the density (blue line with circles). From $N=1000\sim 5000$ the linewidth is not obviously changed. In contrast, by keeping the same density ($n_0k_0^{-3}=0.0037$ for $N$=3000) while varying the OD (magenta line with triangles), the linewidth keeps changing. (b) The frequency shift is calculated for different number of atoms. By keeping constant density (magenta line with triangles), the frequency shift remains almost constant, while for constant OD (blue line with circles), there is a significant variation of  the frequency shift with  $N$. Here the cloud aspect ratio is $R_x:R_y:R_z=2:2:1$ \cite{bromley2016}.  }\label{fig:scaling}
\end{figure}

The interplay between the imaginary and real part of the dipole-dipole interaction has to be carefully accounted for to compare numerical simulations with experimental measurements by doing finite size scaling.  For typical  computation resources  the numerical solution of Eq. (\ref{eq:steady}) is  limited to $\sim 10^4$ atoms. On the other hand  experiments usually operate with ensembles of  tens of millions of atoms. To theoretically model  these large systems  a proper rescaling in the cloud size is necessary. Equation (\ref{eq:1st}) implies that in order to  characterize the effect of radiative interactions one should aim to match the experimental OD, which scales as $\sim N^{1/2}$. On the other hand, to properly  reproduce  the effect of elastic interactions it is better to match the dimensionless number $n_0k_0^{-3}$, which scales as $N^{1/3}$. Therefore, there exist two different ways of rescaling the cloud size, either by keeping the same OD or the same density. In Fig.~\ref{fig:scaling} we show the effect of finite $N$ for a moderate range of atom numbers. Indeed, except from small deviations seen at very low atom number ($<1000$), the linewidth broadening can be well captured by keeping the OD constant, while  the frequency shift is well described by using a constant density. In contrast, if a constant OD is used to compute the frequency shift, the result would considerablly overestimate the shift,  e.g., by a factor of 10 when the $N$ in numerical simulation is 1/100 the atom number in experiment.    The interplay of multiple scattering and density effects is more prominent for larger  $N$ values. To deal with the OD vs density scaling issues  in comparing with experiments the most appropriate rescaling procedure that we found is the following: when computing   the linewidth or peak intensity, the  theory is rescaled accordingly with the experimental OD. However,  the actual OD value is not  exactly matched to the experimental one but to a slightly modified value, $\widetilde{\rm OD}=\eta {\rm OD}$  to account for density effects \cite{bromley2016}. For a moderate window of ${\rm OD}$  values, for example achieved  experimentally by   letting the cloud expand for different times, $\eta$ should be kept fixed. For the frequency shift the theory should be rescaled according to density.

\subsection{Anisotropic features of scattered light}
\begin{figure}[tb]
\centering
\includegraphics[width=0.35\textwidth]{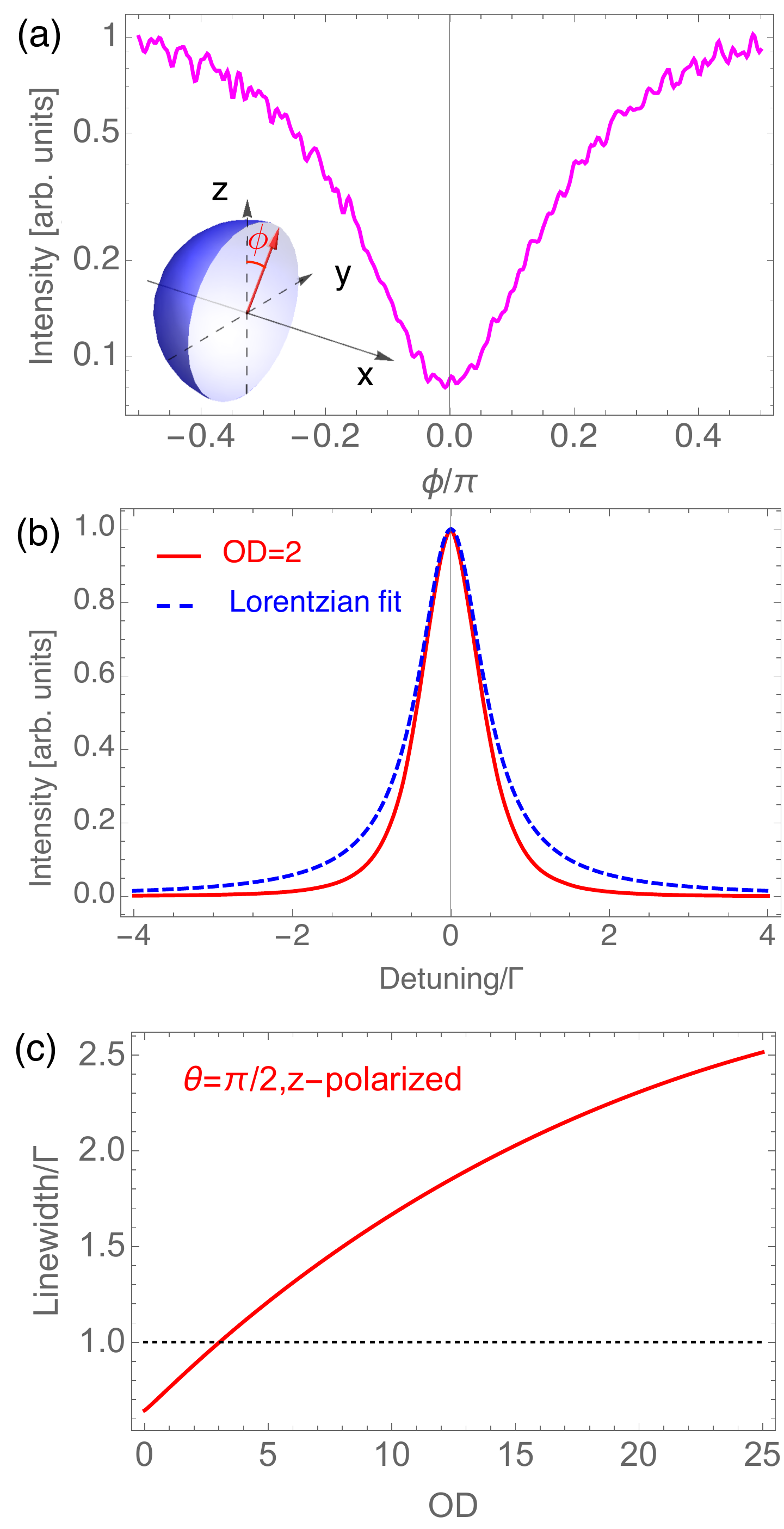}
  \caption{(Color online) {\bf CD Model: effect of laser polarization}. (a) Intensity distribution in the plane perpendicular to the laser propagation direction, {\em i.~e.}, $\theta=\pi/2$ for all $\phi$'s (inset shows the geometry). $\phi=0$ is the z-direction. Here, the incident laser is polarized along z. The intensity detected along the laser polarization  is suppressed compared to other directions. Panels (b)(c) show the lineshape and linewidth of light detected at $\theta=\pi/2,\phi=0$. (b) At small OD, the FWHM linewidth is below the natural linewidth. (c) As OD increases, the linewidth is collectively broadened due to multiple scattering processes (Fig.~\ref{fig:lwcol}(a)). For all panels, degeneracy in excited levels has been assumed. }\label{fig:subrad}
\end{figure}

For independent atoms radiation   along the polarization of the driving laser is forbidden. However,
 dipole-dipole interactions  can generate polarization components different to the driven ones
if the atoms exhibit internal level structure, e.g.,   degenerate  Zeeman levels in the excited state. This is  the case of  a  $J=0\rightarrow J'=1$ transition, where, as  shown  in Fig.~\ref{fig:subrad}, the fluorescence emitted along the laser polarization direction (z-direction, $\theta=\pi/2$ ) is  nonzero. It is, however, much weaker than the intensity emitted along  other directions. On the contrary, for  two-level transitions the polarization of the scattered light is conserved and thus the  emission parallel to the laser polarization is  completely suppressed.  The strong dependence of the  scattered light on polarization and atomic internal structure is most relevant  along the transverse direction. Along  the forward direction  those effects are  irrelevant, as verified by our numerical simulations. From Eq. (\ref{eq:expand}), the lowest order contribution to the intensity detected along the laser polarization direction comes from the first order scattering processes, thus $I\propto \frac{1}{(\Delta^2+\Gamma^2/4)^2}$, which lead to  a ``subradiant'' lineshape ({\em i.~e.} the full width at half maximum (FWHM), $\overline\Gamma_{\rm FWHM}=\sqrt{\sqrt{2}-1}\Gamma<\Gamma,$ is smaller than the one for independent particles). The analytic result  agrees perfectly with the numerical simulation at low OD (Fig.~\ref{fig:subrad}(b)). As the OD increases and interactions become stronger, higher order scattering contributions lead to a collective broadening (linewidth larger  than $\Gamma$) even along this ``single dipole forbidden'' direction, as shown in Fig.~(\ref{fig:subrad} (c)).
\section{Random walk model}
 In this section we use the random walk model to investigate the  role of incoherent scattering processes in collective emission.
 We  focus on the low-intensity regime. Classically, light transport in a disordered medium can be described by a sequence of random scattering events experienced by a photon (see Fig.~\ref{fig:scheme}(a)) \cite{mcrobin1,mcrobin2,resonantmultiple}. The expected number of scattering events is   roughly given by $( {\rm OD}_p)^2$, where  
${\rm OD}_p$ is the peak optical depth, which depends on detuning $\Delta$ as $3N/[k_0^2R^2(1+4\Delta^2/\Gamma^2)]$ .  For simplicity here we also  assume  a spherical cloud. The transmission of light is  given by   $e^{-{\rm OD}_p}$ \cite{fnote2,coherentmultiple}. For the $J=0\rightarrow J'=1$ transition (degenerated $J'=1$ states), the  differential scattering cross section that defines a scattering event is  given by \cite{mcrobin1,resonantmultiple}
\begin{eqnarray}\frac{d\sigma}{d\Omega}({\bf k}_{\rm in},\boldsymbol {\epsilon}_{\rm in}\rightarrow{\bf k}_{\rm out},\boldsymbol {\epsilon}_{\rm out})&=&\frac{3\sigma_0}{8\pi}|\boldsymbol {\epsilon}_{\rm in}^*\cdot\boldsymbol{\epsilon}_{\rm out}|^2,\label{eq:diff}
\end{eqnarray} where ${\bf k}_{\rm in, out}$ is the incident/scattered wavevector, $\boldsymbol{\epsilon}_{\rm in, out}$ is  the polarization of the incident/scattered photon~\cite{mcrobin1,resonantmultiple,mcpol1} and $\sigma_0=3\lambda^2/[2\pi(1+4\Delta^2/\Gamma^2)]$, with $\lambda$ the wavelength of the driving laser.

\begin{figure}
\includegraphics[width=0.45\textwidth]{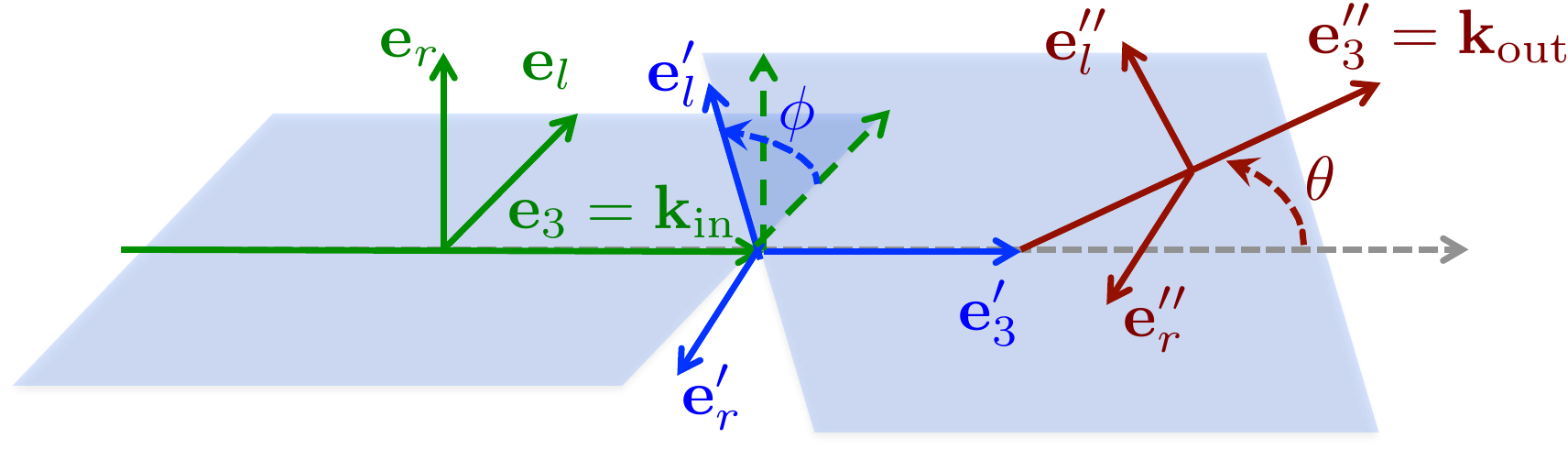}
  \caption{(Color online) \textbf{RW: transformation of Stokes vectors}. 
In the random walk model, a scattering event is determined by two consecutive transformations of  local coordinates: $\{{\bf e}_{\rm 3},{\bf e}_{\rm r},{\bf e}_{\rm l}\}\rightarrow\{{\bf e}_{\rm 3}',{\bf e}_{\rm r}',{\bf e}_{\rm l}'\}$ via rotation $\phi$, and $\{{\bf e}_{\rm 3}',{\bf e}_{\rm r}',{\bf e}_{\rm l}'\}\rightarrow\{{\bf e}_{\rm 3}'',{\bf e}_{\rm r}'',{\bf e}_{\rm l}''\}$ via rotation $\theta$ \cite{mc}.
}\label{fig:stokes}
\end{figure}

To simulate the polarization dependent scattering events as dictated by Eq.~(\ref{eq:diff}), it is convenient to use the Stokes-Mueller formalism~\cite{lighttransfer}. A photon in a given state of polarization can be described by a Stokes vector~\cite{pollightbook}
\begin{eqnarray}
{\bf S}&=&\left(\begin{array}{lcr}
S_0\\
S_1\\
S_2\\
S_3
\end{array}\right)=\left({\begin{array}{c}
\overline{|E_l|^2}+\overline{|E_r|^2}\\
\overline{|E_l|^2}-\overline{|E_r|^2}\\
\overline{E_l^*E_r}+\overline{E_lE_r^*}\\
i(\overline{E_l^*E_r}-\overline{E_lE_r^*})
\end{array}}\right),
\end{eqnarray}
where $E_l$, $E_r$ are the electric field  components projected onto the two orthorgonal axis $\hat{e}_l$ and $\hat{e}_r$ in the plane perpendicular to the wavevector ${\bf k}$. For example, ${\bf S}=(1,1,0,0)$ represents a photon linearly polarized along the reference axis $\hat{e}_l$. A scattering event ${\bf k}_{\rm in},\boldsymbol {\epsilon}_{\rm in}\rightarrow{\bf k}_{\rm out},\boldsymbol {\epsilon}_{\rm out}$ can be determined by two angles: $\theta$ and $\phi$ [see Fig.~\ref{fig:stokes}]. The change of polarization can be obtained from the transformation  ${\bf S}''=M(\theta)\mathcal{R}(\phi){\bf S}^{\rm in}$, where   ${\bf S}^{\rm in}$ is the incident Stokes vector, ${\bf S}''$ is defined with respect to the axis $\hat{e}_l''$, $\hat{e}_r''$ and $\hat{e}_3''$,  and  then ${\bf S}^{\rm out}=\mathcal{R}(\psi){\bf S}''$ transforming back to the original frame $\hat{e}_l$, $\hat{e}_r$ and $\hat{e}_3$~\cite{mc}, with ${\bf S}^{\rm out}$  the scattered Stokes vectors. The scattering matrix that we use,  $M$,  is the scattering matrix that describes Rayleigh scattering~\cite{lighttransfer,resonantmultiple}. It is given by
\begin{eqnarray}
{M(\theta)}&=&\frac{3}{4}\begin{pmatrix}
{\rm cos}^2\theta+1&{\rm cos}^2\theta-1&0&0\\
{\rm cos}^2\theta-1&{\rm cos}^2\theta+1&0&0\\
0&0&2{\rm cos}\theta&0\\
0&0&0&2{\rm cos}\theta
\end{pmatrix}.
\end{eqnarray} The matrix, $\mathcal{R}(\phi)$, is the rotation matrix that rotates the incident axis $\hat{e}_r$ to $\hat{e}_r'$ (perpendicular to the scattering plane),

\begin{eqnarray}
{\mathcal{R}(\phi)}&=&\begin{pmatrix}
1&0&0&0\\
0&{\rm cos}(2\phi)&{\rm sin}(2\phi)&0\\

0&-{\rm sin}(2\phi)&{\rm cos}(2\phi)&0\\
0&0&0&1
\end{pmatrix}
\end{eqnarray} and $\mathcal{R}(\psi)$ is the rotation matrix that transforms the coordinate system $\hat{e}_l''$, $\hat{e}_r''$ and $\hat{e}_3''$ back to the original coordinate system  $\hat{e}_l$,$\hat{e}_r$ and $\hat{e}_3$, and can be found from $\theta$ and $\phi$.
 The polarization of the photons is encoded in the Stokes vector. The probability of a scattering event can be directly calculated from $S''_0/S^{\rm in}_0$. Complete trajectories of the photons can be found from Monte Carlo sampling of  scattering events. As the phase information of photon is not kept in this approach, it is more suitable for describing classical media or hot atoms where phase coherence is not important.

\begin{figure}
\centering
\includegraphics[width=0.5\textwidth]{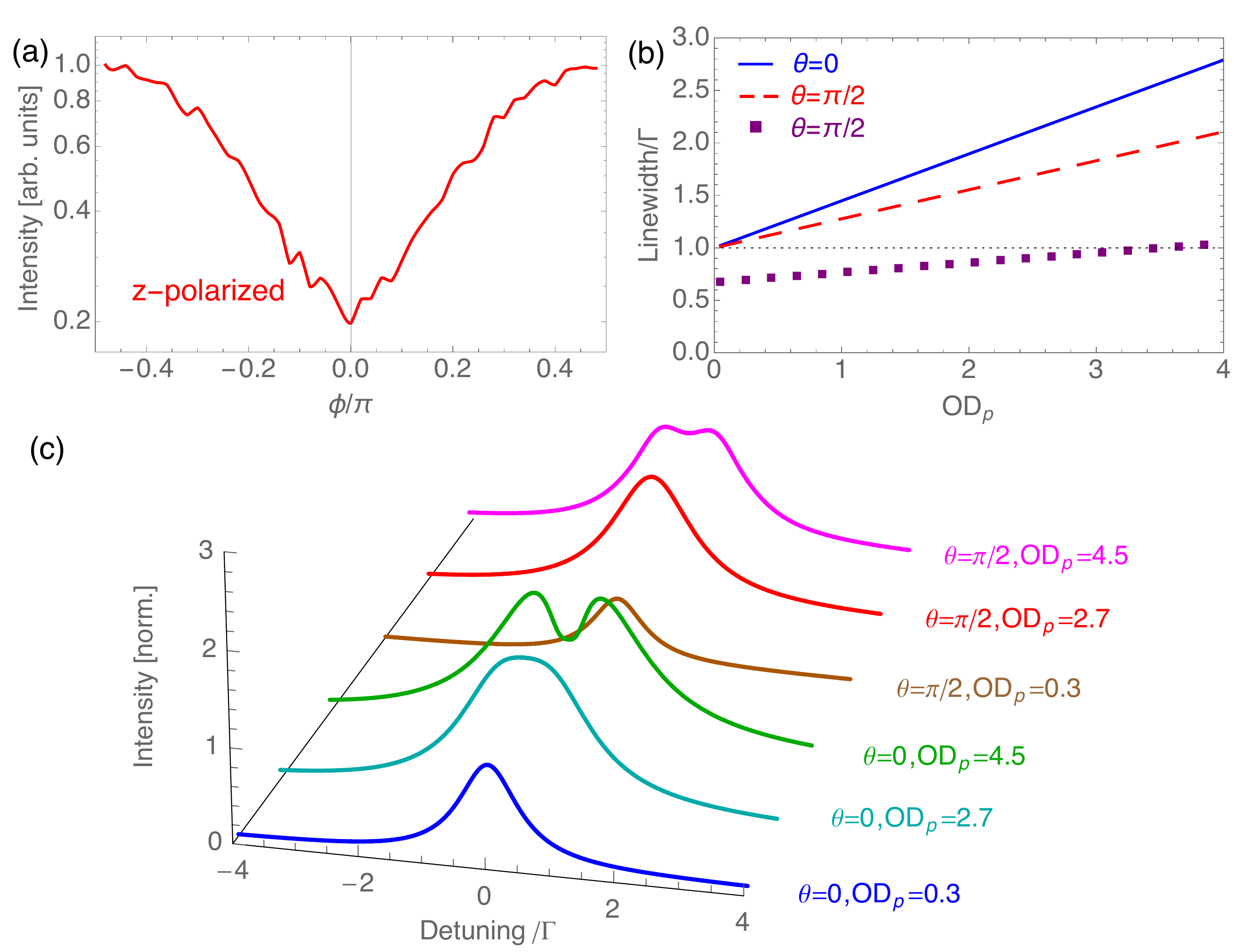}
  \caption{(Color online) {\bf RW Model}. (a) Distribution of on-resonance intensity in the plane $\theta=\pi/2$. Along the direction where single scattering events are forbidden ($\phi=0$), the light intensity is suppressed.  (b) The FWHM linewdith increases linearly with OD$_p$ in a dilute medium.  The linewidth along the direction parallel to the laser polarization can drop  below $\Gamma$ for small (purple squares). (c) Lineshapes for different  OD$_p$s and angles of observation.  The lineshape is Lorentzian for  low OD$_p$ but gets distorted and develops a double-peak structure as the OD$_p$ increases. Here, for all angles, the signal is collected within a small angular region $\delta\theta=5^\circ$. The intensity is normalized to the value at zero detuning for $\theta=0$ and OD$_p$=0.3, and the width of laser beam is 5 times  the Gaussian width of the atomic cloud. Here all OD$_p$ are labeled as the value at $\Delta=0$.}\label{fig:rwlw}
\end{figure}

The polarization of the incident photon is randomized after multiple scattering events. Since
 the intensity detected along the polarization direction of the incident photon  requires at least two scattering events,
   it is suppressed compared to other  directions  but nonzero (Fig.~\ref{fig:rwlw}(a)) and
 the linewidth at small ${\rm OD}_p$ also drops below the natural linewidth (Fig.~\ref{fig:rwlw}(c)) along this  direction. For the other directions,  the intensity distribution is roughly homogeneous and does not exhibit  the collective enhancement along the forward direction observed in the coherent dipole model. The FWHM linewidth for different directions increases linearly with OD up to a moderate value of OD (see Fig.~\ref{fig:rwlw}(b)), and displays a polarization dependence similar to the prediction of  CD. Under this classical treatment  more scattering  processes are expected to  occur with  increasing ${\rm OD}_p$ and those processes  tend to inhibit the transmission of light. As the scattering becomes more frequent, forward scattering decreases and more light is scattered backwards  \cite{labeyrie2004multiple}. Since ${\rm OD}_p$  is maximum  at resonance, $\Delta=0$,  the linewidth develops a  a ``double-peak'' profile as the medium becomes denser (Fig.~\ref{fig:rwlw}(c)). Before the distortion in lineshape develops, the FWHM linewidth also linearly increases with ${\rm OD}_p$. We note that a  similar ``double-peak'' structure also appears in the coherent dipole model, but the physical origin of it is different. In the latter, it only happens at specific small angles where the interference  and multiple scattering effects compete with each other (see Fig.~\ref{fig:lwcol}(c)), and never happens along the forward direction, where the interference effect dominates or the transverse direction, where the multiple scattering effect dominates. In summary, despite of the fact that the RW model does not include coherent emission mechanisms,  it is able to reproduce the collective broadening observed with increasing optical depth  and the sub-natural linewidth present in the direction parallel to the laser polarization at relatively small optical depths. The RW model on the other hand  ignores coherent elastic dipolar interactions and thus does not predict a density shift.

\section{Role of atomic motion}
An assumption made in Sec. \ref{sec:CD} is that the position of the  atoms remains fixed. This assumption is only valid when  atoms move at a rate  slower than the radiative decay rate. When motion is significantly  faster, e.g. hot atomic clouds, the coherence during radiation is smeared out and classical approaches, such as the  RW model, are usually satisfactory \cite{scatreview1999,scatterbook} to describe collective light emission. However,  many   experiments operate in an intermediate regime where coherences cannot be totally disregarded and
  it is not a good approximation to treat atoms as frozen. For example, for the $^1{\rm S}_0\rightarrow {}^3$P$_1$ transition of $^{88}$Sr atoms, the Doppler boadening at $\sim 1~\mu$K, $\Delta_{\rm D}\approx 6\Gamma$, and recoil frequency $\omega_r=\hbar k_0^2/2m\approx 0.6\Gamma$, both comparable to the natural linewidth \cite{Ido2005,Burnett1992}. Consequently for a proper description of light scattering  one needs to account for both photon coherences and atomic motion on an equal footing. Below we present two approximate ways to accomplish this task:

\subsection{Modified frozen dipole model}\label{subsec:modified}

In this section, we discuss a simple way to include the effects of atomic motion on light scattering via a modified CD model. Here we start by  deriving the fluorescence intensity emitted by a weakly driven atom, using a full quantum approach with motional effect included. We consider the states including at most one excitation, and label the relevant quantum states by $\ket{{\bf p}\alpha,0},\ket{{\bf p}g,{\bf k}}$, with $\alpha=\{e,g\}$, {\bf p}  the momentum of the atom,  and ${\bf k}$ is the momentum of the photon in vacuum ($\ket{{\bf p}\alpha,0}$ stands for no photon). For generality we assume here that two counter propagating lasers are used to drive the atoms, carrying  momentum ${\bf k}_0$ and $-{\bf k}_0$ respectivelly. The Hamiltonian is \cite{fnote3}
\begin{eqnarray}
\hat{H}&=&\frac{\hat {\bf p}^2}{2m}+\hbar\omega_a \hat{b}^\dagger\hat{b}+\hbar\Omega[{\rm cos}({\bf k}_0\cdot{\bf r})(e^{-i\omega_Lt}\hat b^\dagger+\mathrm{H.c.})]\nonumber\\&&+\sum_{{\bf k},\boldsymbol{\epsilon}}\hbar\omega_k\hat{a}_{{\bf k}\boldsymbol{\epsilon}}^\dagger\hat a_{{\bf k},\boldsymbol{\epsilon}}-\hbar \sum_{{\bf k},\boldsymbol{\epsilon}}({\bf d}\cdot\hat{\boldsymbol{\epsilon}}_{\bf k}) g_k[e^{i{\bf k}\cdot{\bf r}}\hat a_{{\bf k}\boldsymbol{\epsilon}}\hat b^\dagger+\mathrm{H.c.}].\nonumber\\
\end{eqnarray}
The state vector of the system is
\begin{eqnarray}
\ket{\psi}&=&\sum_\alpha\int d{\bf p}\ket{{\bf p}\alpha,0}A_{\alpha 0}({\bf p},t)e^{-i(E_\alpha+E_p)t/\hbar}\nonumber\\&&+\sum_{{\bf k},\boldsymbol{\epsilon}}\int d{\bf p}\ket{{\bf p}g,{\bf k}}B_{g{\bf k}\boldsymbol{\epsilon}}({\bf p},t)e^{-i(E_g+E_p+\hbar\omega_k)t/\hbar}, \nonumber\\
\end{eqnarray}
where $E_{\alpha}=\omega_a\delta_{\alpha,e}$, $E_p=\frac{p^2}{2m}$, $|A_{\alpha 0}({\bf p},t)|^2$ represents the population in the ground/excited state possessing momentum ${\bf p}$, and $B_{g{\bf k}\boldsymbol{\epsilon}}({\bf p},t)$ is the amplitude of having a photon ${\bf k}$ with polarization $\boldsymbol{\epsilon}$. The state of the system evolves according to
\begin{eqnarray}
i\frac{dA_{g0}({\bf p},t)}{dt}&=&\Omega A_{e0}({\bf p}+{\bf k},t)e^{i(\omega_L-\omega_a)t}e^{-iE_{{\bf p}+{\bf k}_0,{\bf p}}t/\hbar}\nonumber\\&&+\Omega A_{e0}({\bf p}-{\bf k},t)e^{i(\omega_L-\omega_a)t}e^{-iE_{{\bf p}-{\bf k}_0,{\bf p}}t/\hbar},\nonumber\\ \\
i\frac{dA_{e0}({\bf p},t)}{dt}&=&-\sum_{{\bf k},\boldsymbol{\epsilon}}({\bf d}\cdot\hat{\boldsymbol{\epsilon}}_{\bf k})g_kB_{g{\bf k}\boldsymbol{\epsilon}}({\bf p}-{\bf k},t)\nonumber\\&&\times e^{-i(\omega_k-\omega_a)t}e^{-iE_{{\bf p}-{\bf k},{\bf p}}t/\hbar}\nonumber\\&&+\Omega A_{g0}({\bf p}-{\bf k},t)e^{-i(\omega_L-\omega_a)t}e^{-iE_{{\bf p}-{\bf k}_0,{\bf p}}t/\hbar}\nonumber\\&&+\Omega A_{g0}({\bf p}+{\bf k},t)e^{-i(\omega_L-\omega_a)t}e^{-iE_{{\bf p}+{\bf k}_0,{\bf p}}t/\hbar},\nonumber\\
\label{eq:2nd}\\
i\frac{dB_{g{\bf k}\boldsymbol{\epsilon}}({\bf p},t)}{dt}&=&B_{g{\bf k}\boldsymbol{\epsilon}}({\bf p},t)-({\bf d}\cdot\hat{\boldsymbol{\epsilon}}_{\bf k})g_kA_{e0}({\bf p}+{\bf k},t)\nonumber\\&&\times e^{i(\omega_k-\omega_a)t}e^{-iE_{{\bf p}+{\bf k},p}t/\hbar},
\end{eqnarray}
where $E_{p_1,p_2}=\frac{p_1^2-p_2^2}{2m}$. The first term   in Eq. (\ref{eq:2nd}) describes the effect of vacuum photons, which according to Wigner-Weisskopf approach leads to the spontaneous decay with rate $\Gamma$, and can be rewritten as \cite{berman1976}
\begin{eqnarray}
i\frac{dA_{e0}({\bf p},t)}{dt}&=&-i\frac{\Gamma}{2}A_{e0}({\bf p},t)+\Omega A_{g0}({\bf p}-{\bf k},t)\nonumber\\&&\times e^{-i(\omega_L-\omega_a)t}e^{-iE_{{\bf p}-{\bf k}_0,{\bf p}}t/\hbar}\nonumber\\&&+\Omega A_{g0}({\bf p}+{\bf k},t)e^{-i(\omega_L-\omega_a)t}e^{-iE_{{\bf p}+{\bf k}_0,{\bf p}}t/\hbar}.\nonumber\\
\end{eqnarray}
Consider the initial condition $B_{g{\bf k}\boldsymbol{\epsilon}}({\bf p},0)=0$, $A_{g0}=\delta({\bf p}-{\bf p}_0)$, where ${\bf p}_0$ is the initial momentum of the atom. The steady state solution is
\begin{eqnarray}
A_{e0}({\bf p},\infty)=\frac{\Omega[\delta({\bf p}_0-{\bf p}+{\bf k}_0)+\delta({\bf p}_0-{\bf p}-{\bf k}_0)]}{\Delta_{L}+E_{{\bf p}_0,{\bf p}}+i\frac{\Gamma}{2}},\nonumber\\
\end{eqnarray}
with $\Delta_L=\omega_L-\omega_a$. Thus the atomic excitation $\mathcal{A}_e=\int d{\bf p}|A_{e0}({\bf p},\infty)|^2$ indicates two Lorentzian with FWHM=$\Gamma$ and centered at $\omega_r\pm\frac{{\bf k}_0\cdot{\bf p}_0}{m}$.
The  photon emission rate  along a given direction ${{\bf k}_s}$ is
\begin{eqnarray}
I_{{\bf k}_s}=\frac{V\hbar c}{(2\pi)^3}\!\int\! dk ~k^3\sum_{\boldsymbol{\epsilon}}\int \! d{\bf p}~ \lim_{t\to\infty} \frac{|B_{g{\bf k}\boldsymbol{\epsilon}}({\bf p},t)|^2}{t},\nonumber\\
\end{eqnarray}
with ${\bf k}=k\hat{\bf k}_s$. We consider the  transverse intensity  case where, ${\bf k}\cdot{\bf k}_0=0$, and for atomic transitions $\omega_r, ~{\bf p}_0\cdot{\bf k}_0/m\ll\omega_L$, then
\begin{eqnarray}
I_{{\bf k}_s}&\approx&\frac{\omega_L^4}{c^3}\frac{ d^2\Omega^2}{8\pi^2\varepsilon_0}[\frac{1}{(\Delta-\omega_r+\frac{{\bf k}_0\cdot{\bf p}_0}{m})^2+\frac{\Gamma^2}{4}}\nonumber\\&&+\frac{1}{(\Delta-\omega_r-\frac{{\bf k}_0\cdot{\bf p}_0}{m})^2+\frac{\Gamma^2}{4}}],
\end{eqnarray} The emitted light  intensity exhibits  the same profile as the atomic excitation $\mathcal{A}_{e}$.

For a single atom, to leading order, motion modifies the emitted light intensity by adding two natural corrections: a Doppler shift $\propto
\frac{{\bf k}_0\cdot{\bf p}_0}{m}$, a velocity dependent  modification of the effective laser detuning experienced by an atom, and a recoil shift, $\omega_r$, which physically accounts for the fact that to compensate for the energy imparted to the atom via photon recoil,  the incident laser needs to have a higher frequency to be resonant with the atomic transition.

\begin{figure}
\centering
\includegraphics[width=0.35\textwidth]{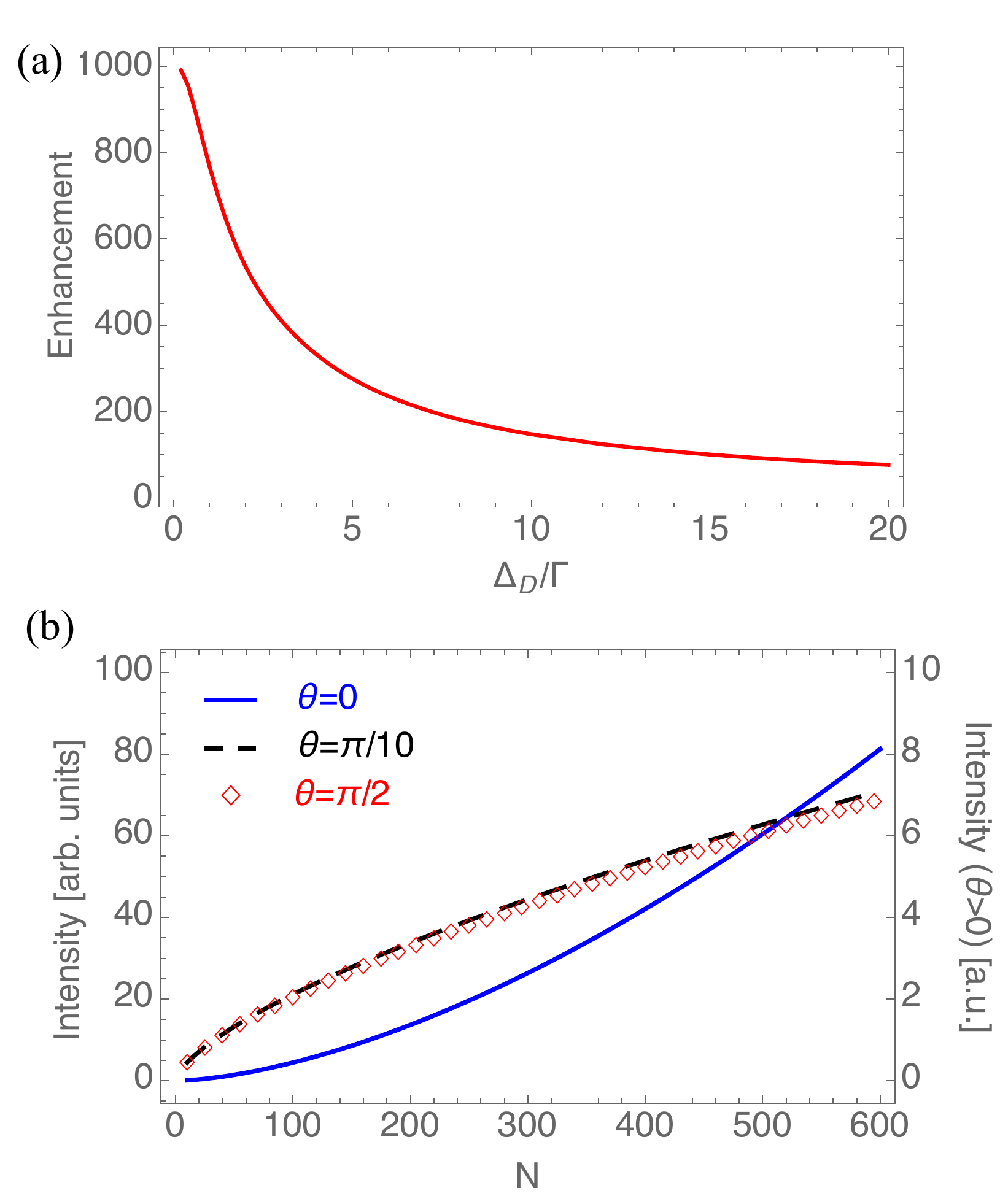}
  \caption{(Color online) {\bf Modified frozen dipole model: peak intensity}. (a) Collective forward enhancement:  forward  intensity normalized by the transverse intensity as a function of Doppler width.  (b)  Intensity detected along different directions at Doppler width $\Delta_{\rm D}=6\Gamma$ as a function of atom number for fixed cloud size (when $N=500$, OD=2,$n_0k_0^{-3}$=0.0015). $I\sim N^{1.6}$ for $\theta=0$, and $I\sim N^{0.7}$ for $\theta=\pi/2$. The right vertical axis labels the intensity for $\theta=\pi/2,\pi/10$.}\label{fig:dopeff}
\end{figure}

We generalize the above results to deal with  motional effects in the  many-body system by  introducing
 random detunings $\delta \nu$ for each atom, and  by sampling them  according to a Maxwell-Boltzmann distribution $P(\delta\nu)=\frac{1}{\sqrt{2\pi}\widetilde \Delta_{\rm D}}{\rm exp}\left(-\frac{\delta \nu^2}{2 \widetilde\Delta_{\rm D}^2}\right)$ that accounts for the Doppler shifts \cite{juhashift}. Here $\widetilde\Delta_{\rm D}=\Delta_{\rm D}/\sqrt{8{\rm ln}2}$.  We denote this approximation as the modified frozen dipole model.
The random detunings have two straightforward effects: (i) dephasing and (ii) reduced light  scattering probability \cite{bromley2016}.
(i) Dephasing reduces the  forward enhancement of fluorescence intensity as shown in Fig. \ref{fig:dopeff} (a).
This can be understood  even at the level of  non-interacting two-level atoms, where Doppler shifts  modify  atomic coherences as (for simplicity we assume a single beam illumination):
\begin{eqnarray}
b_j&=&\frac{\Omega e^{i{\bf k}_0\cdot {\bf r}_j}}{(\Delta-\delta\nu_j)+i\Gamma/2}.\label{eq:RCD}
\end{eqnarray} Using  Eq. (\ref{eq:RCD}), the intensity  along a generic direction becomes
\begin{eqnarray}
I_{\rm incoh}&=&\frac{1}{\sqrt{2\pi}\widetilde\Delta_{\rm D}}\int d\delta\nu_j |b_j|^2e^{-\delta\nu_j^2/2\widetilde\Delta_{\rm D}^2}.
\end{eqnarray}
 On the other hand, the coherent scattering in the forward direction which takes into account pairwise atomic contributions, becomes
\begin{eqnarray}
I_{\rm coh}\!&=&\!\frac{1}{2\pi\widetilde\Delta_{\rm D}^2}\!\!\int\!\! d\delta\nu_{j}d\delta\nu_{j'} b_{j}b^*_{j'}e^{-\delta\nu_{j}^2/2\widetilde\Delta_{\rm D}^2}e^{-\delta\nu_{j'}^2/2\widetilde\Delta_{\rm D}^2}.
\end{eqnarray}
 The  on-resonance enhancement factor is thus
\begin{eqnarray}
\frac{I_{\rm coh}}{I_{\rm incoh}}&=&\frac{\sqrt{\frac{\pi}{2}}e^{\frac{1}{8\widetilde\Delta_{\rm D}^2/\Gamma^2}}{\rm Erfc}(\frac{1}{2\sqrt{2}\widetilde\Delta_{\rm D}/\Gamma})}{2\widetilde\Delta_{\rm D}/\Gamma},
\end{eqnarray} where ${\rm Erfc}$ is the complementary error function. This shows an exponential suppression of the forward interference that depends on $\widetilde\Delta_{\rm D}/\Gamma$.

The reduced light  scattering probability, on the other hand, competes with dephasing since it suppresses multiple scattering, and as a consequence promotes collective  enhancement. Effectively, it brings the system closer to the small OD regime (Fig.~\ref{fig:multipscat}(b)). The motion induced suppression of multi-scattering processes is also signaled in the frequency shift (Fig.~\ref{fig:shifdens}) \cite{juhashift,antoine2016}, which keeps increasing until a larger value of density in the presence of motion.
\subsection{Semi-classical approach}

Laser light mediated forces on atoms are a fundamental concept in atomic physics and lay the foundations of  laser trapping and cooling \cite{coolingreview}. They can be accounted  for at the semiclassical level by explicitly including the  position ${\bf r}_i$ and the momentum ${\bf p}_i$ degrees of freedom of the atoms, and solving for their dynamics   while feeding those back into the quantum dynamics of the  internal degrees of freedom. An explicit description of this procedure is presented below.

 \begin{figure}
\centering
 \includegraphics[width=0.3\textwidth]{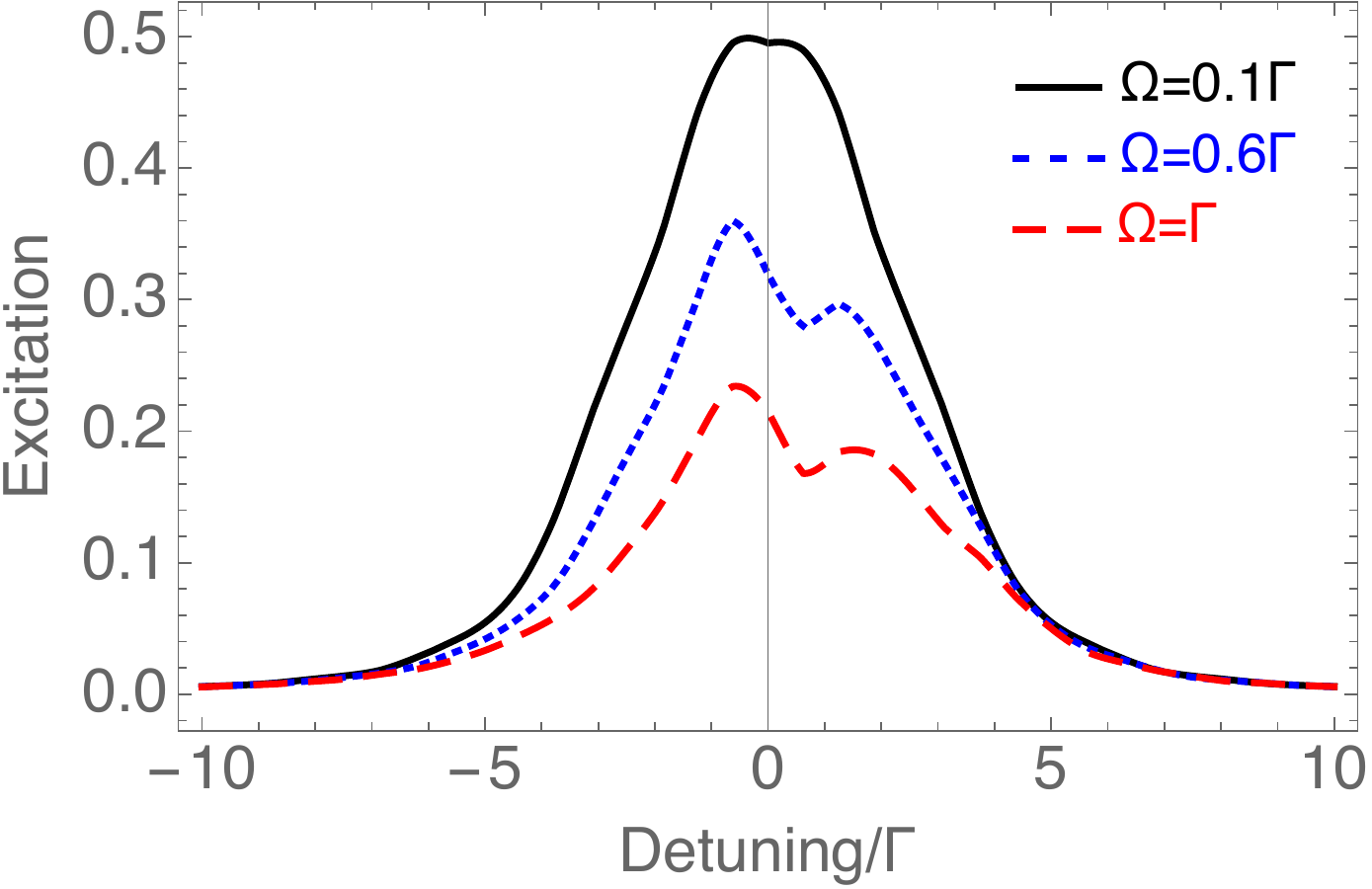}
   \caption{(Color online) {\bf Semi-classical model: single-atom lineshape}. The excitation lineshape is calculated for three different driving strength, each normalized by $\Omega^2$. The atomic motion is allowed to be one-dimensional and parallel to the laser propagating direction. Here we used $\omega_r=0.6\Gamma$ (the value of the $^1 {\rm S}_0\rightarrow {}^3$P$_1$ transition of Sr \cite{Ido2005}). }\label{fig:dali}
 \end{figure}

For simplicity we will  assume a two-level transition, with $\omega_a=k_0 c$ the frequency of transition and ${\bf d}$ the dipole moment. This  condition is achievable in experiments,  for example, by  applying a large magnetic field to split apart ($|\Delta^\alpha-\Delta^\gamma| \gg {\mathcal G}\Gamma$) the excited state levels and thus  energetically suppressing population of the ones not directly driven by the laser.

Let us first ignore the driving term and include motion in the  dipolar coupling after eliminating the electromagnetic vacuum modes. The Hamiltonian, including the atoms and free-space electromagnetic field is
\begin{eqnarray}
H&=&\hbar\omega_a\sum_j\hat{b}_j^\dagger\hat{b}_j+\sum_{{\bf k},\boldsymbol{\epsilon}}\omega_ k\hat{a}^\dagger_{{\bf k}\boldsymbol{\epsilon}}\hat{a}_{{\bf k}\boldsymbol{\epsilon}}-\sum_j\sum_{{\bf k},\boldsymbol{\epsilon}}g_k\nonumber\\&&\times({\bf d}\cdot\hat{\boldsymbol{\epsilon}}_{\bf k})[(e^{i{\bf k}\cdot {\bf r}_j}\hat{a}_{{\bf k}\boldsymbol{\epsilon}}+\mathrm{H.c.})(\hat{b}_j^\dagger+\mathrm{H.c.})].
\end{eqnarray}
The evolution of atomic dipoles and the field modes are
\begin{eqnarray}
\frac{d\hat{a}_{{\bf k}\boldsymbol{\epsilon}}}{dt}&=&-i\omega_k\hat{a}_{{\bf k}\boldsymbol{\epsilon}}+\frac{i}{\hbar}\sum_jg_k({\bf d}_j\cdot\boldsymbol{\epsilon}_{\bf k})[e^{-i{\bf k}\cdot{\bf r}_j}(\hat{b}_j+\mathrm{H.c.})],\label{eq:a}\nonumber\\
\\
\frac{d\hat{b}_{j}}{dt}&=&-i\omega_a\hat{b}_j+i\sum_{{\bf k},\boldsymbol{\epsilon}}\frac{g_k}{\hbar}({\bf d}_j\cdot\hat{\boldsymbol{\epsilon}}_{\bf k})[e^{i{\bf k}\cdot{\bf r}_j}\hat{a}_{{\bf k}\boldsymbol{\epsilon}}\hat{s}_j+\mathrm{H.c.}],\label{eq:b}\nonumber\\
\\
\frac{d\hat{s}_{j}}{dt}&=&-i\sum_{{\bf k},\boldsymbol{\epsilon}}\frac{g_k}{\hbar}({\bf d}_j\cdot\hat{\boldsymbol{\epsilon}}_{\bf k})[e^{i{\bf k}\cdot{\bf r}_j}\hat{a}_{{\bf k}\boldsymbol{\epsilon}}\hat{b}_j^\dagger-e^{-i{\bf k}\cdot{\bf r}_j}\hat{a}_{{\bf k}\boldsymbol{\epsilon}}^\dagger\hat{b}_j\nonumber\\&&-e^{i{\bf k}\cdot{\bf r}_j}\hat{a}_{{\bf k}\boldsymbol{\epsilon}}\hat{b}_j+e^{-i{\bf k}\cdot{\bf r}_j}\hat{a}_{{\bf k}\boldsymbol{\epsilon}}^\dagger\hat{b}_j^\dagger],\label{eq:z}
\end{eqnarray}
where $\hat{s}_j=\hat{b}_j^\dagger\hat{b}_j-\hat{b}_j\hat{b}_j^\dagger$, and $s_j=\langle \hat{s}_j\rangle$ gives the inversion of the $j^{\rm th}$ atom. We have assumed that  internal operators commute with external operators, and neglected the diffusion of the atomic wavepacket.  Eq. (\ref{eq:a}) can be formally integrated to obtain
\begin{eqnarray}
\hat{a}_{{\bf k}\boldsymbol{\epsilon}}(t)&=&\hat{a}_{{\bf k}\boldsymbol{\epsilon}}(0)-i\sum_j\frac{g_k}{\hbar}({\bf d}_j\cdot\hat{\boldsymbol{\epsilon}}_{\bf k})\int dt'~e^{i{\bf k}\cdot{\bf r}_j+i\omega_k(t'-t)}\nonumber\\&&\times (\hat{b}_j+\mathrm{H.c.}),\label{eq:a1}
\end{eqnarray}
in which we have assumed  the external motion is much slower than the internal dynamics so that ${\bf r}_j(t')\approx{\bf r}_j(t)$, and the interaction between atoms and the field modes is weak so that $\hat{b}_j(t')\approx\hat{b}_j(t)e^{-i\omega_a(t'-t)}$. Substituting Eq. (\ref{eq:a1}) into Eq. (\ref{eq:b}), we obtain the equation for the quantum averaged quantities
\begin{eqnarray}
\frac{db_{j}}{dt}&=&-i\omega_ab_j+s_j\sum_{l}\sum_{{\bf k},\boldsymbol{\epsilon}}\frac{g_k^2}{\hbar^2}({\bf d}_j\cdot\hat{\boldsymbol{\epsilon}}_{\bf k})({\bf d}_l\cdot\hat{\boldsymbol{\epsilon}}_{\bf k})\nonumber\\&&\times \{e^{i{\bf k}\cdot{\bf r}_{jl}}[b_l^*(\pi\delta(\omega_k+\omega_a)-i{\rm P}\frac{1}{\omega_k+\omega_a})\nonumber\\&&+b_l(\pi\delta(\omega_k-\omega_a)-i{\rm P}\frac{1}{\omega_k-\omega_a})]\nonumber\\&&-e^{-i{\bf k}\cdot{\bf r}_{jl}}[b_l^*(\pi\delta(\omega_k-\omega_a)+i{\rm P}\frac{1}{\omega_k-\omega_a})\nonumber\\&&+b_l(\pi\delta(\omega_k+\omega_a)+i{\rm P}\frac{1}{\omega_k+\omega_a})]\},
\end{eqnarray} 
where 
$\mathcal{O}=\langle\hat{\mathcal{O}}\rangle$ for any atomic operator $\hat{\mathcal{O}}$, and  we have utilized the fact that $\langle \hat{a}_{{\bf k}\boldsymbol{\epsilon}}(0)\rangle=0$, and assumed that atomic motion is classical \cite{smith91,smith92}. Changing $\sum_{\bf k}=\frac{V}{(2\pi)^3}\int d\Omega\, dk~k^2$, and applying rotating wave approximation, we have
\begin{eqnarray}
\frac{db_{j}}{dt}&=&-i\omega_ab_j+i\sum_{l\neq j}s_jb_l(g_{jl}-if_{jl})-\frac{\Gamma}{2}b_j,
\end{eqnarray} with   $f(0)=\Gamma$. The equation for $s_j$ can be derived in a similar way and is given by 
\begin{eqnarray}
\frac{ds_j}{dt}&=&-\Gamma s_j-2i\sum_{l\neq j}(g_{jl}-if_{jl})b_j^*b_l+2i\sum_{l\neq j}(g_{jl}+if_{jl})b_l^*b_j,\nonumber\\
\end{eqnarray}
 and
\begin{eqnarray}
g(r)&=&-\frac{3\Gamma}{4}(z_1(\theta)\frac{{\rm cos}k_0r}{k_0r}+z_2(\theta)[\frac{{\rm cos}k_0r}{k_0^3r^3}+\frac{{\rm sin}k_0r}{k_0^2r^2}]),\nonumber\\ \\
f(r)&=&\frac{3\Gamma}{4}(z_1(\theta)\frac{{\rm sin}k_0r}{k_0r}+z_2(\theta)[\frac{{\rm sin}k_0r}{k_0^3r^3}\!-\!\frac{{\rm cos}k_0r}{k_0^2r^2}]),
\end{eqnarray} where $z_1(\theta)={\rm sin}^2\theta$, and $z_2(\theta)=(3{\rm cos}^2\theta-1)$.
For the momentum,
\begin{eqnarray}
\frac{d\hat{\bf p}_j}{dt}&=&-\nabla \hat{H}\nonumber\\
&=&-\sum_{{\bf k},\boldsymbol{\epsilon}}g_k({\bf d}_j\cdot\boldsymbol{\epsilon}_{\bf k})(i{\bf k}e^{i{\bf k}\cdot{\bf r}_j}\hat{a}_{{\bf k}\boldsymbol{\epsilon}}\hat{b}_j^\dagger-i{\bf k}e^{-i{\bf k}\cdot{\bf r}_j}\hat{a}_{{\bf k}\boldsymbol{\epsilon}}^\dagger\hat{b}_j\nonumber\\&&
+i{\bf k}e^{i{\bf k}\cdot{\bf r}_j}\hat{a}_{{\bf k}\boldsymbol{\epsilon}}\hat{b}_j-i{\bf k}e^{-i{\bf k}\cdot{\bf r}_j}\hat{a}_{{\bf k}\boldsymbol{\epsilon}}^\dagger \hat{b}_j^\dagger).
\end{eqnarray} After substituting Eq. (\ref{eq:a1}), taking the quantum average and performing a similar integration procedure of as above, we obtain
\begin{eqnarray}
\frac{d {\bf p}_j}{dt}&=&\sum_{l\neq j}[\mathpzc{g}_{jl}(b_jb_l^*+\mathrm{H.c.})-i\mathpzc{f}_{jl}(b_jb_l^*-\mathrm{H.c.})],
\end{eqnarray} 
with $\mathpzc{g}_{jl}=-\nabla g_{jl}$ and $\mathpzc{f}_{jl}=-\nabla f_{jl}$. As the dispersive force $\mathpzc{g}_{jl}$ is a steep function of $r_{jl}$,  it dominates at short distance, and atoms are drastically accelerated/decelerated. Both the dispersive and dissipative forces are anisotropic and  couple motion along  different directions.

The prior equations of motion however are not the full story.
 Due to the presence of spontaneous emission and radiative interactions, the atomic momentum diffuses over time, which can be described by including classical noise $d\xi_i^\alpha$ in the equation of motion for $ p_i^\alpha$. The components of these noises are correlated, and characterized by the  diffusion matrix
\begin{eqnarray}
E[ d\xi_i^\alpha(t) d\xi_i^\beta(t')]&=&\delta_{\alpha,\beta}\frac{2-\delta_{\alpha,z}}{20}\hbar^2k_0^2\Gamma (s_i+1)\delta(t-t'),\nonumber\\\\
E[ d\xi_i^\alpha (t)d\xi_j^\beta(t')]&=&-\hbar^2k_0^2\nabla_\alpha\nabla_\beta f(r_{ij})\Re[b^*_ib_j]\delta(t-t'),\nonumber\\
\end{eqnarray}
where $E[\cdot]$ denotes the expectation value.

The momentum diffusion matrix can be found from
\begin{eqnarray}
\mathcal{D}_{jl}&=&\frac{d\langle\hat{\bf p}_j\hat{\bf p}_l\rangle}{dt}-\langle {\bf p}_j\rangle\frac{d\langle {\bf p}_l\rangle}{dt}-\frac{d\langle {\bf p}_j\rangle}{dt}\langle{\bf p}_l\rangle\\
&=&\sum_{{\bf k}_1,\boldsymbol{\epsilon_1}}\sum_{{\bf k}_2,\boldsymbol{\epsilon}_2}g_{k_1}g_{k_2}({\bf d}_j\cdot\boldsymbol{\epsilon}_1)({\bf d}_l\cdot\boldsymbol{\epsilon}_2)(i{\bf k}_1e^{i{\bf k}_1\cdot{\bf r}_j}\hat{a}_{{\bf k}_1\boldsymbol{\epsilon}_1}\hat{b}_j^\dagger\nonumber\\&&-i{\bf k}_1e^{-i{\bf k}_1\cdot{\bf r}_j}\hat{a}_{{\bf k}_1\boldsymbol{\epsilon}_1}^\dagger\hat{b}_j\nonumber\\&&
+i{\bf k}_1e^{i{\bf k}_1\cdot{\bf r}_j}\hat{a}_{{\bf k}_1\boldsymbol{\epsilon}_1}\hat{b}_j-i{\bf k}_1e^{-i{\bf k}_1\cdot{\bf r}_j}\hat{a}_{{\bf k}_1\boldsymbol{\epsilon}_1}^\dagger \hat{b}_j^\dagger)\nonumber\\&&\times (i{\bf k}_2e^{i{\bf k}_2\cdot{\bf r}_l}\hat{a}_{{\bf k}_2\boldsymbol{\epsilon}_2}\hat{b}_l^\dagger-i{\bf k}_2e^{-i{\bf k}_2\cdot{\bf r}_l}\hat{a}_{{\bf k}_2\boldsymbol{\epsilon}_2}^\dagger\hat{b}_l\nonumber\\&&
+i{\bf k}_2e^{i{\bf k}_2\cdot{\bf r}_l}\hat{a}_{{\bf k}_2\boldsymbol{\epsilon}_2}\hat{b}_l-i{\bf k}_2e^{-i{\bf k}_2\cdot{\bf r}_l}\hat{a}_{{\bf k}_2\boldsymbol{\epsilon}_2}^\dagger \hat{b}_l^\dagger)\nonumber\\
&=&-\hbar^2k_0^2\nabla\nabla f_{ij}\Re[b_j^*b_l].
\end{eqnarray}

In dense clouds momentum diffusion from radiative interactions can give rise at long times to significant heating. This heating  was reported to be one of the main limiting mechanisms  in laser cooling \cite{dalidop,smith92}. At short times, $t \Gamma\sim 1$, with low densities and weak probes, $\Omega<\Gamma$, the momentum diffusion is not  prominent, and since this is the regime we are interested in this work, we will ignore momentum diffusion in our  calculations.

 For driving the atoms  we will consider the case of two counter-propagating lasers with wavevector $\pm{\bf k}_0$, propagating along $z$. The laser generates an additional force $-2\hbar {\bf k}_0\Omega  {\rm sin}({\bf k}_0\cdot {\bf r}_i)\Re\{ b_i(t)\}$ and drives the laser coherences as $\frac{db_{i}}{dt}=i\Omega\cos({\bf k}_0\cdot{\bf r}_i)s_i$ \cite{coherentmultiple}. Those terms need to be added in the corresponding equations.

From previous sections, it has been  shown that except for a narrow cone around the forward direction, the fluorescence intensity is $I({\bf r}_s)\propto\sum_i s_i$. Since we expect motion to further reduce the effect of coherence,  we will focus on  the transverse scattering, and compute  $\sum_i s_i$. 
Fig. \ref{fig:dali} shows the lineshape of a single atom for different $\Omega$ calculated from the semi-classical approach, without accounting for momentum diffusion. We have verified in our numerical simulations that it can be safely ignored for most of the parameters presented there.

As a result of  atomic motion, the laser-atom system is not in a  stationary state. We show the results for driving time $ t=5/\Gamma$.  When $\Omega\ll \Gamma$, the lineshape is a Voigt profile with the Doppler width determined by the velocity. When $\Omega\sim \Gamma$, there is a distortion in the lineshape, with more intensity at $\Delta<0$. This is because for $\Delta<0$ the laser force decelerates the atom, while for $\Delta>0$ the atom is heated up. With reduced/increased velocity, the atom is on average more/less excited, resulting in a distorted lineshape. The center of the line is therefore shifted to the red.

\begin{figure}
\centering
\includegraphics[width=0.35\textwidth]{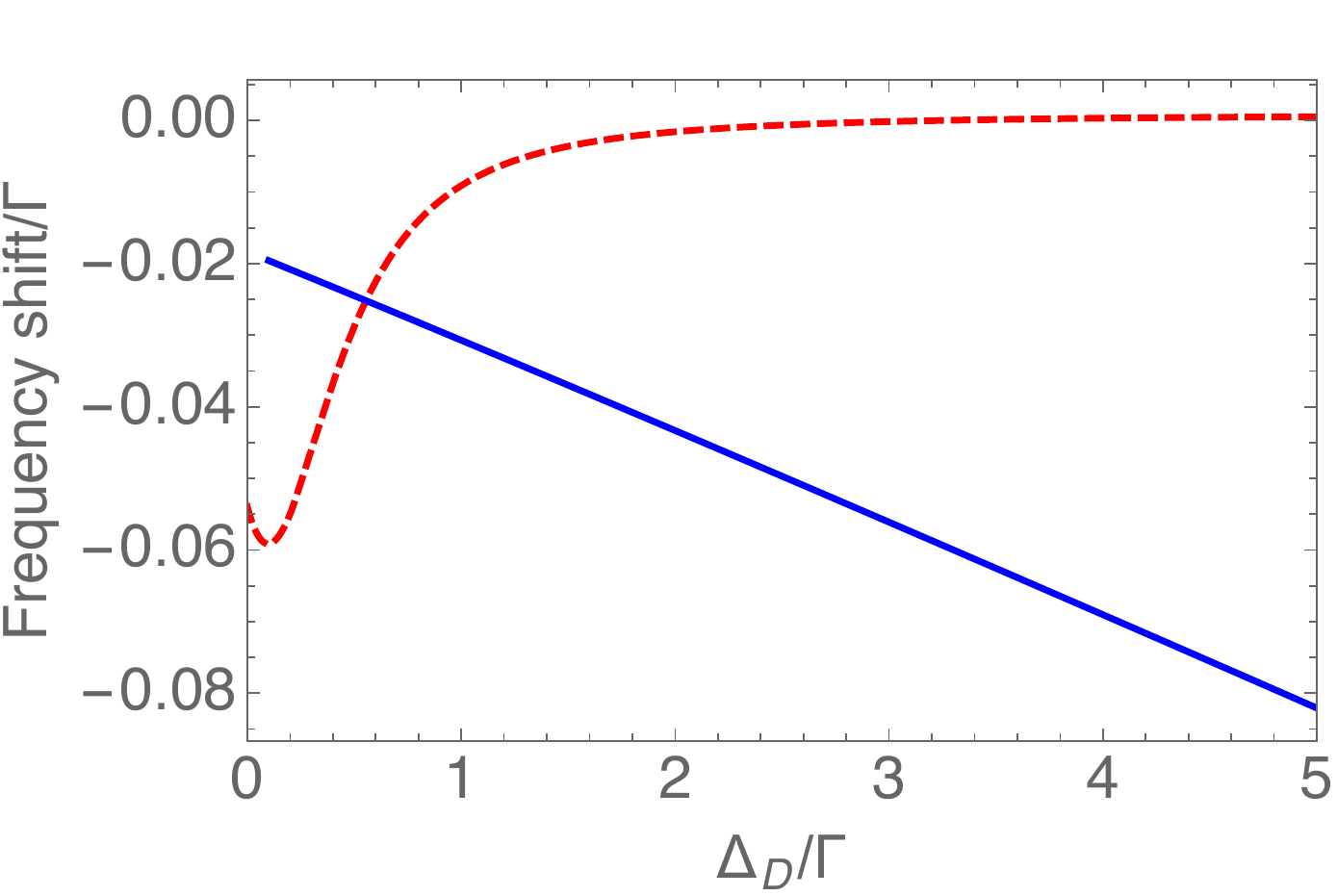}
  \caption{(Color online) {\bf Semi-classical model: effect of motion on frequency shift}. Two atoms  are driven by a pair of counter-propagating  lasers, with $\Omega=0.1\Gamma$, $\omega_r=0.6\Gamma$. The red dashed line shows the result of semi-classical model where motion is allowed in three dimensions. The blue line shows the result from the modified CD. The center of excitation lineshape is calculated at $t=5/\Gamma$ for different Doppler width, for both models. 
}\label{fig:shiftT}
\end{figure}

Motion can  significantly  modify   the interactions between atoms \cite{collmotion}. We study the
effect of motion on the frequency shift  in Fig.~\ref{fig:shiftT}. Dipolar induced frequency shifts were previously discussed in  Sec. \ref{sec:CD}.  Since we focus on low driving fields, again  we neglect  momentum diffusion in these calculations. The calculations  show that unless $\Delta_{\rm D}$ is very small, atomic motion leads to a fast suppression of the  frequency shift. 
Only when $\Delta_{\rm D}\ll \Gamma$, the frequency can be increased by motion and this is the regime where the modified frozen dipole model is qualitatively valid; recall, it  predicts always an increase of density shift with Doppler broadening.    We note that at $\Delta_{\rm D}\rightarrow 0$ the frequency shift obtained using  the modified frozen dipole model,  is slightly smaller than that one obtained from the  semi-classical approach. This is a consequence of the distortion caused by laser cooling/heating which  additionally shifts the spectral  line. For $\Delta_{\rm D}\gtrsim \Gamma$ motion needs to be properly accounted for and the modified frozen dipole model is not reliable.


\section{Conclusion}
 We theoretically studied the propagation of light through a cold atomic medium. We presented two different microscopic models, the ``coherent dipole model'' and the ``random walk model'', and  analyzed how
  the light polarization, optical depth  and density
affect the  linewidth broadening, intensity and line center shift  of the emitted light. We  showed that the  random walk model, which neglects photonic phase coherence,  can fairly capture the collective broadening (narrowing) of the emission linewidth but on the other hand does not predict a density shift.
 Due to the limitation of computation capacity, the numerical simulation of CD is usually restricted to $\sim 10^4$ atoms, which is much smaller than that in some cold atom experiments \cite{subradiance2015,haveypra2013}. Nevertheless, the understanding of the underlying physics allowed us to perform an appropriate rescaling in the cloud size which we used  to compare with experiments \cite{bromley2016}.
 We further developed generalized models that explicitly take into account motional effects. We showed that atomic motion can lead to drastic dephasing and reduction in the collective effects, together with a distortion in the lineshape. While the modified frozen dipole model predicts a monotonic increase of the density shift with increasing motion, the semiclassical model, which properly accounts for recoil effects, predicts that this behavior only holds at slow motion $\Delta_{\rm D}\ll \Gamma$. Instead, as atoms move faster, motional effects  start to become dominant, the cloud expands  and the frequency shift decreases.
None of the  presented theoretical models, however, can explain the large density shift measured  in the $^1{\rm S}_0\rightarrow{}^3{\rm P}_1$ transition of $^{88}$Sr atoms \cite{Ido2005}. It will be intriguing to determine what are the actual physical processes that cause this large shift.

\section{ Acknowledgements}
 The authors wish to acknowledge useful discussions with JILA Ye group,  Alexey Gorshkov,  Juha Javanainen, Mark Havey,  Muray Holland, Michael L. Wall, Jose D'Incao, Matthew Davis, and Johannes Schachenmayer. We would like to especially thank Robin Kaiser for helpful discussions and insightful comments.  This work was supported by the NSF (PIF-1211914 and
PFC-1125844), AFOSR, AFOSR-MURI, NIST and ARO. Computations utilized the Janus supercomputer, supported by NSF (award number CNS-0821794), NCAR, and CU Boulder/Denver.

\appendix
\section{Optical depth of a  cloud with Gaussian distribution}\label{app:od}
We consider an atomic cloud with  a Gaussian distribution $n(x,y,z)=n_0e^{-\frac{x^2}{2R_x^2}-\frac{y^2}{2R_y^2}-\frac{z^2}{2R_z^2}}$, where $n_0$ satisfies $\int dxdydz\, n(x,y,z)=N$, and $N$ is the total number of atoms. Along the line of observation, {\em e.g.} $\hat{x}$, the on resonance optical depth is related to the resonant scattering cross section, which for  the $J=0\rightarrow J=1$ transition is $\sigma_{sc}=\sigma_0(\Delta=0)=\frac{6\pi}{k_0^2}$,  and the column density averaged over the profile perpendicular to this direction~\cite{dalibard,subradiance},
\begin{eqnarray}
{\rm OD}&=&[\int dy dz \,n(y,z)]^{-1}\int dy dz \, n(y,z) {\rm OD}(y,z)\nonumber\\
&=&[\int dy dz \,n(y,z)]^{-1}\int dy dz \,n(y,z)\int dx n(x,y,z)\sigma_{sc}\nonumber\\
&=&[\int dy dz \,n(y,z)]^{-1} dy dz\, n(y,z)e^{-\frac{y^2}{2R_y^2}-\frac{z^2}{2R_z^2}}\nonumber\\&&\times\int dx\, n_0e^{-\frac{x^2}{2R_x^2}}\sigma_{sc}\nonumber\\
&=&\frac{3N}{2k_0^2R_yR_z}\nonumber\\
&=&\frac{3N}{2k_0^2R_{\perp}^2}.
\end{eqnarray}
With laser detuning $\Delta$, the optical depth is ${\rm OD}/(1+4\Delta^2/\Gamma^2)$.

\bibliographystyle{apsrev}
\bibliography{refscat}

\end{document}